\documentclass[a4paper,fleqn,usenatbib]{mnras}

\usepackage[T1]{fontenc}
\usepackage{ae,aecompl}

\usepackage{graphicx}	
\usepackage{amsmath}	
\usepackage{amssymb}	
\usepackage{rotating}
\usepackage{natbib}
\usepackage{color}
\usepackage{soul}
\usepackage[normalem]{ulem}

\usepackage{array,multirow}

\def \arcsec {\hbox{$^{\prime\prime}$}}

\def\phn{\phantom{0}}

\newcommand{\degree}{\hbox{$^\circ$}}

\newcommand{\perbeam}{beam$^{-1}$}

\def \vhel{\ifmmode{~V_{{\rm HEL}}}\else{~$V_{{\rm HEL}}$}\fi}
\def \vsys{\ifmmode{~V_{{\rm SYS}}}\else{~$V_{{\rm SYS}}$}\fi}
\def \HA {\ifmmode{{\rm\H}\alpha}\else{${\rm\ H}\alpha$}\fi}
\def \farcs{\hbox{$~\!\!^{\prime\prime}$}} 

\def \msun{\ifmmode{{\rm\ M}_\odot}\else{${\rm\ M}_\odot$}\fi}

\def \myr{\ifmmode{{\rm\ M}_\odot{\rm\ yr}^{-1}}
        \else{${\rm\ M}_\odot$ yr$^{-1}$}\fi}

\def \mdot{\ifmmode{\dot{M}}\else{$\dot{M}$}\fi}
\def \tena#1 #2 {\ifmmode{#1 \times 10^{#2}}\else{$#1 \times 10^{#2}$}\fi}
\def \kms{\ifmmode{~{\rm km\,s}^{-1}}\else{~km s$^{-1}$}\fi}

\def \apj{ApJ}
\def \mnras{MNRAS}
\def \pasp{PASP}

\def \aap{A\&A}

\def \apjs{ApJS}
\def \apjl{ApJL}
\def \nat{Nat}
\def \iaucirc{IAUC.}

\def \skytel{Sky and Telescope}
\def \nar{NewAR}

\newcommand{\xtenineteen}{XTE\,J1908$+$094}
\newcommand{\grs}{GRS\,1915$+$105}
\newcommand{\gro}{GRO\,J1655$-$40}

\definecolor{mynotecol}{RGB}{0,127,0}

\title[Jet-ISM interactions in \xtenineteen]{Resolved, expanding jets in the Galactic black hole candidate \xtenineteen}

\author[A. P. Rushton et al.]{A.~P. Rushton,$^{1,2}$\thanks{E-mail: anthony.rushton@physics.ox.ac.uk (APR)}
J.~C.~A. Miller-Jones,$^3$
P.~A. Curran,$^{3}$\thanks{Deceased}
G.~R. Sivakoff,$^4$
\and M.~P. Rupen,$^5$
Z. Paragi,$^6$
R.~E. Spencer,$^{7}$
J. Yang,$^{8,9}$
D. Altamirano,$^2$
\and T. Belloni,$^{10}$
R.~P. Fender,$^1$
H.~A. Krimm,$^{11,12}$
D. Maitra,$^{13}$
S. Migliari,$^{14,15}$
\and
D.~M. Russell,$^{16}$
T.~D. Russell,$^{17,3}$
R. Soria$^{3,18}$ and 
V. Tudose$^{19}$
\\
$^{1}$ Department of Physics, Astrophysics, University of Oxford, Keble Road, Oxford OX1 3RH, UK\\
$^{2}$ School of Physics and Astronomy, University of Southampton, Highfield, Southampton SO17 1BJ, UK\\
$^{3}$ International Centre for Radio Astronomy Research -- Curtin University, GPO Box U1987, Perth, WA 6845, Australia\\
$^{4}$ Department of Physics, University of Alberta, 4-181 CCIS, Edmonton, AB T6G 2E1, Canada\\
$^{5}$ National Research Council of Canada, Herzberg Astronomy and Astrophysics, Dominion Radio Astrophysical Observatory, \\PO Box 248, Penticton, BC V2A 6J9, Canada\\
$^{6}$ Joint Institute for VLBI ERIC (JIVE), Postbus 2, NL-7990 AA Dwingeloo, the Netherlands\\
$^{7}$ Jodrell Bank Observatory, University of Manchester, Macclesfield, Cheshire SK11 9DL, UK\\
$^{8}$ Department of Earth and Space Sciences, Chalmers University of Technology, Onsala Space Observatory, SE-439 92 Onsala, Sweden\\
$^{9}$ Shanghai Astronomical Observatory, Chinese Academy of Sciences, 200030 Shanghai, P. R. China\\
$^{10}$ Istituto Nazionale di Astrofisica, Osservatorio Astronomico di Brera, Via E. Bianchi 46, I-23807 Merate, Italy\\
$^{11}$ Universities Space Research Association, 7178 Columbia Gateway Dr, Columbia, MD 21046, USA\\
$^{12}$ National Science Foundation, 4201 Wilson Blvd, Arlington, VA 22230, USA\\
$^{13}$ Department of Physics \& Astronomy, Wheaton College, Norton, MA 02766, USA\\
$^{14}$ European Space Astronomy Centre (ESAC), 28692 Villanueva de la Ca\~{n}ada, Madrid, Spain\\
$^{15}$ Department of Quantum Physics and Astrophysics \& Institute of Cosmos Sciences, University of Barcelona, 08028 Barcelona, Spain\\
$^{16}$ New York University Abu Dhabi, PO Box 129188, Abu Dhabi, UAE\\
$^{17}$ Anton Pannekoek Institute for Astronomy, University of Amsterdam, PO Box 94249, NL-1090 GE Amsterdam, the Netherlands\\
$^{18}$ Sydney Institute for Astronomy, School of Physics A28, The University of Sydney, NSW 2006, Australia\\
$^{19}$ Institute for Space Sciences, Atomistilor 409, PO Box MG-23, 077125 Bucharest-Magurele, Romania
}

\begin{document}

\date{Accepted 2017 February 27. Received 2017 February 26 ; in original form 2017 January 16}

\pagerange{\pageref{firstpage}--\pageref{lastpage}} \pubyear{2017}

\maketitle

\label{firstpage}

\begin{abstract}
Black hole X-ray binaries undergo occasional outbursts caused by changing inner accretion flows. Here we report high-angular resolution radio observations of the 2013 outburst of the black hole candidate X-ray binary system \xtenineteen, using data from the VLBA and EVN. We show that following a hard-to-soft state transition, we detect moving jet knots that appear asymmetric in morphology and brightness, and expand to become laterally resolved as they move away from the core, along an axis aligned approximately $-11$\degree\ east of north. We initially see only the southern component, whose evolution gives rise to a 15-mJy radio flare and generates the observed radio polarization.  This fades and becomes resolved out after 4 days, after which a second component appears to the north, moving in the opposite direction. From the timing of the appearance of the knots relative to the X-ray state transition, a 90\degree\ swing of the inferred magnetic field orientation, the asymmetric appearance of the knots, their complex and evolving morphology, and their low speeds, we interpret the knots as working surfaces where the jets impact the surrounding medium.  This would imply a substantially denser environment surrounding \xtenineteen\ than has been inferred to exist around the microquasar sources \grs\ and \gro.
\end{abstract}



\section{Introduction}

Galactic black hole X-ray binaries (BH XRBs) exhibit a plethora of
accretion `states' \citep[e.g.][]{2010LNP...794...53B}, allowing us to study the different regimes 
in which a black hole accretes matter and interacts with the
surrounding environment. Over relatively short time scales, the 
spectral energy distribution of a BH XRB can change significantly due to variations in the structure and geometry of the accretion flow. BH XRBs also drive extended 
outflows (or jets), whose structure appears to be directly related to these changes in
the inner accretion flow. During accretion states that are dominated 
by hard, non-thermal X-ray emission, an AU-scale, compact, quasi-steady
jet can be observed, whereas transitions from the hard state to softer X-ray 
states are associated with the ejection of relativistically-moving `knots' of plasma that are no longer causally connected to the 
accretion flow \citep[as reviewed in, e.g.][]{2006csxs.book..381F}.  While the
structure and composition of the jets are not known, \citet{2005Natur.436..819G}, \citet{2007MNRAS.376.1341R} and \citet{2015MNRAS.446.3579S} have shown that the power of the outflows can be comparable to the total bolometric luminosity of the binary
system.

Perhaps the most famous jet-producing black hole is \grs,
which was the first Galactic source observed to show superluminal
motion \citep[e.g.][]{1994Natur.371...46M,1999MNRAS.304..865F}. The
source has shown repeated major radio outbursts, in which discrete knots of optically-thin plasma are directly resolved by high angular resolution monitoring whenever a radio flare occurs during the plateau state
\citep{2010MNRAS.401.2611R}. Although
transient radio ejecta have been best studied in \grs, there are a handful
of other systems that have shown discrete, resolved jet knots on milliarcsecond scales, including GRO J1655$-$40 \citep{1995Natur.374..141T,1995Natur.375..464H}, Cyg X$-$3 \citep{2001ApJ...553..766M,2004ApJ...600..368M}, H1743$-$322 \citep{2012MNRAS.421..468M}, and XTE\,J1752$-$223 \citep{Yang10}. These sources all appear to display resolved jets following a hard-to-soft X-ray state change, with typical proper
motions of 10--100 mas d$^{-1}$ implying velocities $v\sim0.1-0.9$c \citep[for distances of 1--10 kpc][]{2009ApJ...695.1111L,2014ApJ...796....2R}. Some of these sources have shown evidence
for deceleration of the knots
\citep[e.g.][]{2002Sci...298..196C,2005ApJ...632..504C,2011MNRAS.415..306M,2011MNRAS.418L..25Y},
implying that an external shock is produced as the jets interact with
the interstellar medium (ISM) \citep{2002Sci...298..196C}.

Most of the individual jet ejecta from BH XRBs have remained unresolved perpendicular to the jet axis, showing no lateral expansion \citep{2006MNRAS.367.1432M} or interaction with the
ISM.  One of the few Galactic cases in which the jet/ISM interaction has been directly resolved is the (persistent) neutron star XRB Sco X-1. \citet{2001ApJ...558..283F} analysed the expansion of two radio lobes from Sco X-1, which reached a minimum resolved size of $4\times10^8$~km ($\sim1$~mas) over a period of a few hours. They argued that the lobes must consist of electrons adiabatically expanding in a working surface, akin to the hot spots found in many extragalactic radio sources.  Another case where lateral expansion of the jet ejecta has been directly resolved is the BH XRB XTE J1752$-$223, for which an expansion speed of $0.9\pm0.1$\,mas\,d$^{-1}$ (i.e.\ $0.05c[d/10{\rm\,kpc}]$) was measured in one of the decelerating jet components, again interpreted as interaction with the ISM \citep{Yang10}.

A basic phenomenological explanation for the jet behaviour \citep{2004MNRAS.355.1105F} couples increased mass accretion
rate to an increasingly powerful compact jet, until a rapid increase of the Lorentz factor causes internal shocks to form
within the jets \citep{2000A&A...356..975K}, producing the observed discrete knots of plasma. Models show that shell collisions can give rise to internal shocks, which act as an electron re-energization
mechanism in an adiabatic conical jet
\citep[e.g.][]{2006MNRAS.367.1083K,2010MNRAS.401..394J}. 

While internal shocks can temporarily illuminate jets, more continuous, steady jets in microquasars should eventually terminate in strong shocks and could inflate radio lobes \citep[e.g.][]{1992Natur.358..215M,1992ApJ...401L..15R,2010Natur.466..209P} similar to extragalactic jet sources. However, \citet{2002A&A...388L..40H} argued that the direct detection of radio lobes from X-ray binaries is difficult due to the their low surface brightness, whereas indirect detection (e.g. X-ray hot spots) is common. Based on how far the jet ejecta propagated downstream before decelerating, they estimated the densities of the media surrounding \grs\ and \gro\ to be  $n\leq10^{-3}\,\rm{cm}^{-3}$, implying previous outbursts could have evacuated low density bubbles around these binaries. However, they also argued that hotspots can suddenly appear once the ejecta reach the boundary of the radio lobe and collide with the denser ISM. 

A further complication can arise if the binary has a high velocity due to a natal supernova kick. In this case, the interaction of the jets with the ISM could then lead to the production of asymmetric trails and bow shocks 
\citep{2008ApJ...686.1145H,2009MNRAS.397L...6W,2011ApJ...742...25Y}.

\subsection{XTE J1908$+$094}

\xtenineteen\ was serendipitously discovered by the Proportional Counter Array (PCA) on board the {\it Rossi X-ray Timing Explorer (RXTE)}, and by {\it BeppoSAX}, during an outburst in 2002
\citep{2002IAUC.7856....1W,2002IAUC.7861....2F}.  The X-ray spectral evolution was studied in detail by \citet{2002A&A...394..553I} and \citet{2004ApJ...609..977G}, who found that after a rising hard state phase, the source underwent a hard-to-soft spectral state transition typical of BH XRBs.  It remained in the soft state for 58\,d before transitioning back to the hard state and decaying.  A second X-ray outburst took place approximately a year after the first, peaking in January 2003.

A radio counterpart to the original 2002 outburst
was first discovered in March 2002 by the Very Large Array \citep{2002IAUC.7874....1R}, and
remained active for 2--3 months \citep{2002IAUC.8029....2R}. A near-infrared (NIR) counterpart
was initially reported by \citet{2002MNRAS.337L..23C} using the European Southern Observatory's New Technology Telescope (NTT), and was later resolved into two separate objects
\citep{2006MNRAS.365.1387C}, whose properties were used to conclude that the black hole candidate system is indeed a
low-mass X-ray binary. During the subsequent decay of the outburst, \citet{2004MNRAS.351.1359J}
investigated the disc-jet coupling with (near-)simultaneous radio and X-ray observations of
\xtenineteen\ using the VLA, Westerbork Synthesis Radio Telescope
(WSRT) and {\it Chandra}.

The source distance is poorly constrained, although \citet{2002A&A...394..553I} used the peak bolometric flux of the 2002 outburst to place a lower limit of $>6.8$\,kpc.  Using the measured flux of the soft-to-hard X-ray state transition, which is known to occur at 1--3\% of the Eddington luminosity \citep{Maccarone03,Kalemci13}, \citet{2015MNRAS.451.3975C} found a distance range of 4.8--13.6\,kpc (for black hole masses of $3$--$10M_{\odot}$).  Throughout this paper we therefore assume a canonical distance of 8\,kpc.

A new outburst from \xtenineteen\ was detected in late 2013 by {\it Swift}/BAT \citep{2013ATel.5523....1K}, triggering a multi-wavelength observing campaign. The 15--50\,keV X-ray flux observed
with the {\it Swift}/BAT hard X-ray transient monitor was observed to increase from 2013 October 26 (MJD~56591), reaching $\sim60$ mCrab by October 28. The detection of hard X-rays triggered pointed observations by 
the XRT instrument on board {\it Swift}, which on October 29 (MJD 56594.84) found an unabsorbed flux
of $1.5 \pm 0.1 \times 10^{-9}$~erg s$^{-1}$cm$^{-2}$ in the 0.3--10\,keV band, and a power-law spectrum with $\Gamma = 1.63 \pm 0.07$, confirming that the source was in a hard X-ray spectral state \citep{2013ATel.5529....1K}. The hard X-ray flux measured by {\it Swift}/BAT peaked at $\sim120$~mCrab on MJD\,56595 and subsequently decreased.  The following two {\it Swift}/XRT observations (taken on November 1 and 3; MJD 56597.92 and 56599.53) showed the source to be in an intermediate state, with X-ray spectra still dominated by the power-law component but with an increasing contribution from the disc blackbody.  The X-ray power-law index was also seen to soften over this period, from $\Gamma=1.57$ on October 29 to $\Gamma=2.30$ by November 3 \citep{Zhang15}.  By the time of the fourth {\it Swift}/XRT observation on November 8 (MJD 56604.86), the source was in a soft X-ray spectral state, where it stayed until it became Sun-constrained in early December.  \citet{Zhang15} also found evidence for a non-constant inner disc radius during the soft state, and a disc luminosity that varied with temperature as $L\propto T^2$ rather than the $L\propto T^4$ expected from a standard Shakura-Sunyaev thin disc.  They interpreted this as evidence of an optically-thick, advection-dominated slim disk \citep*[the ``apparently standard" state of][]{Kubota01}.

In addition to the {\it Swift}/XRT observations, both \citet{Zhang15} and \citet{Tao15} analyzed a set of {\it NuSTAR} observations taken on November 8--9 (MJD 56604.76--56605.90) during the soft X-ray spectral state.  They detected a $\sim40$-ks flare during which the X-ray emission increased by up to 40\%, driven by an increase in the power-law component of the spectrum, which also showed a significant spectral softening during the flare. Both works concluded that this could be due to an ejection event, or possibly a change in the properties of the corona.  While the black hole spin could not be constrained from X-ray spectral fitting in the soft state, both groups found similar inclination angles of 20--40\degree\ to the line of sight, which is similar to the $45\pm8$\degree\ inferred by \citet{Miller09} from their modelling of the relativistic disk reflection during a bright hard state.

Radio emission from the 2013 outburst of \xtenineteen\ was first detected by \citet{Miller-Jones13} with the Karl G. Jansky Very Large Array (VLA), and subsequently monitored by the Arcminute
Microkelvin Imager Large Array (AMI-LA) \citep{Rushton13a,Rushton13b} and the Australia Telescope Compact Array \citep[ATCA;][]{Coriat13}. The detectable radio flux lasted for just over 20 days. Initially, the source exhibited quasi-steady, flat spectrum
radio emission at a level of $\sim1$~mJy, simultaneous with the hard X-ray rise.  On November 5 (MJD 56601), AMI-LA detected a rapid radio flare to 15\,mJy, which decayed over the course of a few hours.  The radio emission subsequently quenched to $<0.32$\,mJy on MJD 56602.7, before undergoing a much longer-duration flare that peaked at 13\,mJy on MJD 56607.1 and decayed slowly over the following ten days, becoming undetectable by MJD 56618 \citep{2015MNRAS.451.3975C}.

As BH XRB outbursts are known to be associated with resolved radio jets \citep{2006csxs.book..381F}, we triggered a series of high spatial-resolution Very Long Baseline Interferometry (VLBI) monitoring observations following the detection of the X-ray state transition. In this paper we present these high-resolution observations, taken with the Very Long Baseline Array (VLBA) and the European VLBI Network (EVN) telescopes, which resolved the radio emission into moving, expanding jet knots. We describe the observations in Section \ref{section:observations}, our fitting of the expansion and motion of the resolved jet structure in Section \ref{section:analysis}, and possible interpretations of the observed knots in Section \ref{section:discussion}.

\section{Observations and Reduction}
\label{section:observations}

\begin{table*}
\footnotesize
\caption{Results of VLBI observations. VLBA and EVN observations begin with project codes BM and RR respectively. ${S}$ and ${N}$ denote southern and northern components, respectively, for the one epoch in which both components were detected.}
\begin{center}
\addtolength{\tabcolsep}{-3pt}
\begin{tabular}{lcccccccc}
\hline
Epoch 	&	Start date	&	UT range	&	Project 	&	 Freq.	&	 Beam size 	&	 Peak flux 	&	 Total flux 	&	 RMS noise 	\\
(MJD) 	&	(yyyy-mm-dd)	&	(hh:mm--hh:mm) 	&	code	&	 (GHz)	&	 (mas$\times$mas) 	&	 ($\mu$Jy bm$^{-1}$) 	&	 ($\mu$Jy) 	&	 ($\mu$Jy bm$^{-1}$) 	\\
\hline																	
56603.78	&	 2013-11-07 	&	 18:07 -- 19:11 	&	BM382A 	&	8.3	&	 \phn $3.23\times2.11$ 	&	 \phn $607\pm\phn38$ 	&	 \phn\phn$759\pm\phn77$ 	&	\phn 34	\\
56604.62	&	 2013-11-08 	&	 11:16 -- 18:43 	&	RR007A 	&	4.9	&	 \phn $8.75 \times 6.52$ 	&	 $1290\pm102$ 	&	 \phn$1200\pm130$ 	&	112	\\
56604.82	&	 2013-11-08 	&	 18:36 -- 20:53 	&	BM382B 	&	8.3	&	 \phn $2.71\times1.59$ 	&	 $2342\pm\phn22$ 	&	 \phn$3637\pm\phn95$ 	&	\phn 18	\\
56605.79	&	 2013-11-09 	&	 18:10 -- 19:39 	&	BM382C 	&	8.3	&	 \phn $3.28\times1.93$ 	&	 $1799\pm\phn42$ 	&	 \phn$7898\pm223$ 	&	\phn 31	\\
56606.80	&	 2013-11-10 	&	 17:37 -- 20:39 	&	BM382D 	&	8.3	&	 \phn $3.53\times1.64$ 	&	 $2254\pm\phn36$ 	&	 $15054\pm462$ 	&	\phn 31	\\
56607.78$^{S}$	&	 2013-11-11 	&	 17:37 -- 19:39 	&	BM382E 	&	8.3	&	 \phn $4.01\times1.96$ 	&	 \phn$386\pm\phn44$ 	&	 \phn$2829\pm394$ 	&	\phn 34	\\
56607.78$^{N}$	&		&		&		&		&	  	&	 \phn$921\pm\phn34$ 	&	 \phn$1468\pm\phn82$ 	&	\phn 34	\\
56608.64	&	 2013-11-12 	&	 12:12 -- 18:42 	&	RR007B 	&	4.9	&	 \phn $7.44 \times 6.27$ 	&	 \phn$690\pm\phn\phn8$ 	&	 \phn\phn$940\pm\phn30$ 	&	\phn \phn 8	\\
56609.00	&	 2013-11-13 	&	 22:06 -- 00:54 	&	BM382F 	&	8.3	&	 \phn $3.56\times1.00$\phn 	&	 \phn$680\pm\phn30$ 	&	 \phn$1851\pm631$ 	&	\phn 19	\\
56609.54	&	 2013-11-13 	&	 10:21 -- 12:58 	&	RR007B 	&	4.9	&	 $12.92 \times 5.76$ 	&	 \phn$267\pm\phn\phn9$ 	&	 \phn\phn$544\pm\phn20$ 	&	\phn \phn 9	\\
56610.90	&	 2013-11-14 	&	 20:07 -- 22:53 	&	BM382G 	&	8.3	&	 \phn $2.43\times1.13$ 	&	$<\phn84$	&	N/A	&	\phn 28	\\
56612.92	&	 2013-11-16 	&	 20:38 -- 23:24 	&	BM382H 	&	8.3	&	 \phn $2.50\times0.95$ 	&	$<135$	&	N/A	&	\phn 45	\\
56617.90	&	 2013-11-21 	&    20:09 -- 22:55 	&	BM382I 	&	8.3	&	 \phn $2.29\times0.88$ 	&	$<100$	&	N/A	&	\phn 33	\\
56742.25	&	 2014-05-14 	&	 02:09 -- 09:41 	&	RR009 	&	4.9	&	\phn $5.87\times4.72$ 	&	$<\phn65$	&	N/A	&	\phn 22	\\
\hline
\end{tabular}
\end{center}
\label{table:results}
\end{table*}

\subsection{VLBI observations}

\subsubsection{VLBA}

Following the detection of radio emission from the 2013 outburst, we observed \xtenineteen\ over nine epochs with the Very Long Baseline Array (VLBA), spaced over 14\,d, as detailed in Table~\ref{table:results}.  At each epoch, we observed at a central frequency of 8.4\,GHz, using 256\,MHz of bandwidth.  We used J1800+3848 as a fringe finder source, and the observations were phase-referenced to the nearby (0.39\degree) calibrator source J1907+0907.  The phase referencing cycle time was 3\,min, spending 130\,s on the target and 50\,s on the calibrator in each cycle.  Every seventh cycle we substituted the scan on \xtenineteen\ for a scan on an extragalactic check source, J1905+0952 (0.9\degree\ from both \xtenineteen\ and J1907+0907).  Observation durations ranged from 75\,min to 4\,hr, and for all observations longer than 3.5\,hr, we spent 30\,min at the start and end of the run observing a geodetic block, using multiple calibrator sources spread across the sky to remove residual tropospheric delays and clock errors, to improve the accuracy of our phase referencing.

The data were correlated using the VLBA-DiFX software correlator \citep{2007PASP..119..318D}, and reduced using standard procedures within the Astronomical Image Processing System \citep[\textsc{aips;}][]{2003ASSL..285..109G}.

\subsubsection{EVN}

In October 2013 we made three Target of Opportunity (ToO) observations using the European VLBI Network (EVN) in rapid-response e-VLBI mode \citep{2008evn..confE..40S}, under program codes RR007A and RR007B. As an ad hoc `Out of Session' experiment, only six stations were available for the first observation (On, Jb, Ys, Sh, Hh \& Tr), whereas the remaining two observations were part of a regular e-VLBI session, with ten stations available (Ef, Jb, Mc, Nt, On, Tr, Ys, Wb, Sh \& Hh). As such, although the target was detected in all three observations, it was not possible to unambiguously determine the astrometric position of the first epoch (MJD 56604.6; RR007A) due to its poor \textit{uv}-coverage. All observations were performed with a data transmission rate of 1024\,Mbps using dual polarisation, giving a total bandwidth of 128\,MHz centred at a frequency of 4.9\,GHz and correlated by the EVN Software Correlator at JIVE \citep{2015ExA....39..259K}. The observations were phase referenced to J1905+0952, and we included brief scans on J1907+0907 as a positional check source. The cycle time was 4.5\,min, with each cycle comprising 2\,min on the calibrator and 2.5\,min on the target.

During the first half of 2014 the target appeared to return to the hard state and we requested an additional e-VLBI ToO observation that was taken in May 2014 with the EVN (project code RR009). The EVN did not detect the source during this experiment.

All EVN data calibration was initially performed using the JIVE/EVN pipeline \citep{2002astro.ph..5118R} which uses \textsc{parseltongue} \citep{2006ASPC..351..497K}, a python wrapper for the NRAO Astronomical Image Processing System (\textsc{AIPS}). Standard delay, phase and amplitude solutions were derived from bright standard calibrator sources and interpolated to the two nearby reference sources. Further
time-dependent phase solutions were calculated for J1905+0952, which
in turn were interpolated to J1907+0907 and the target, for
astrometric reference. The four epochs of the target field were edited for bad data points and imaged using a Briggs weighting of 1. As with our VLBA data, an image plane deconvolution of the synthesised beam was performed for each epoch to fit the centroid position and measure any extension of the radio emission.

\subsection{Astrometric correction}\label{alignment}

\begin{figure}
\centering 
\includegraphics[width=7cm,angle=0]{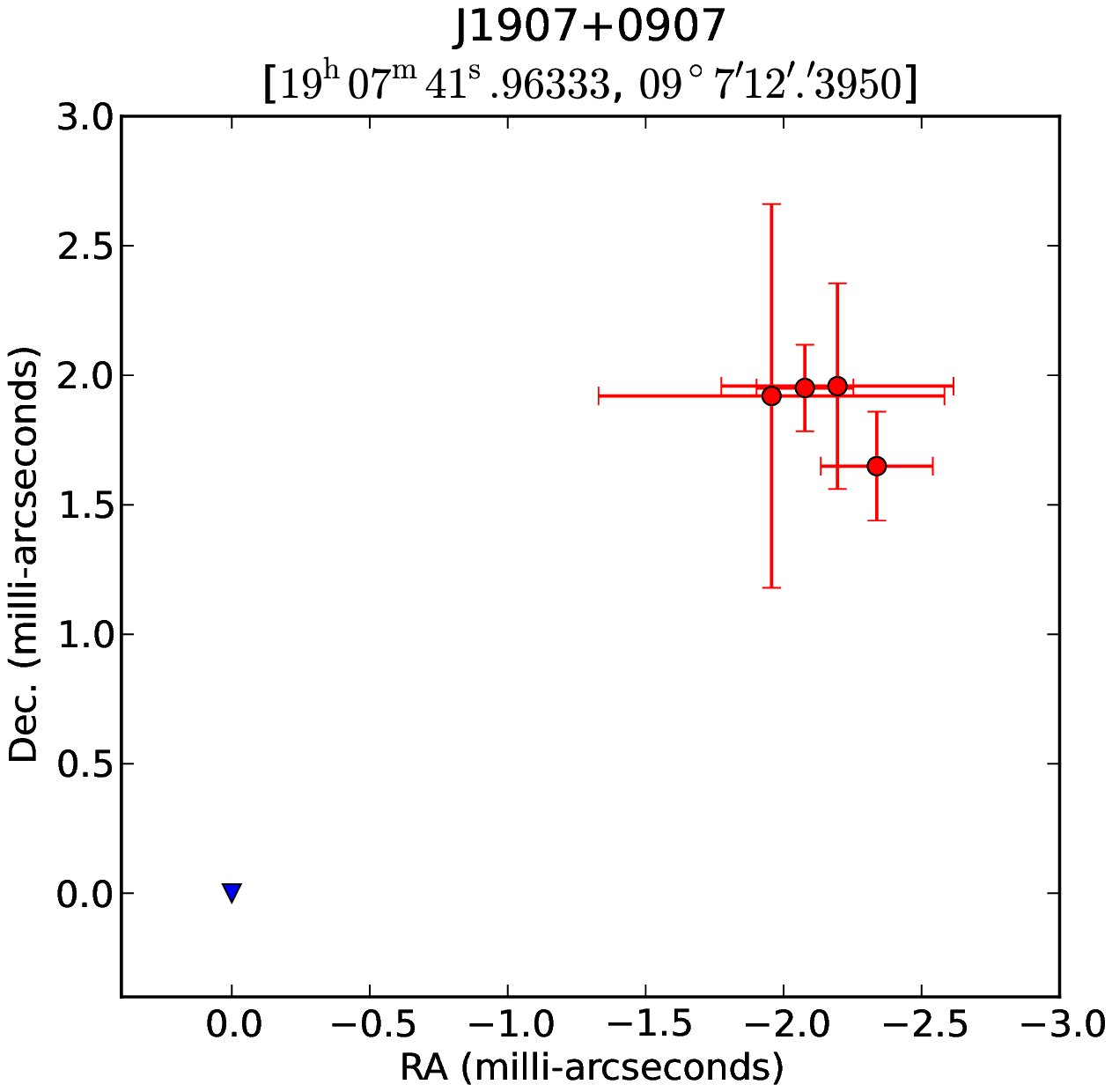} 
\includegraphics[width=7cm,angle=0]{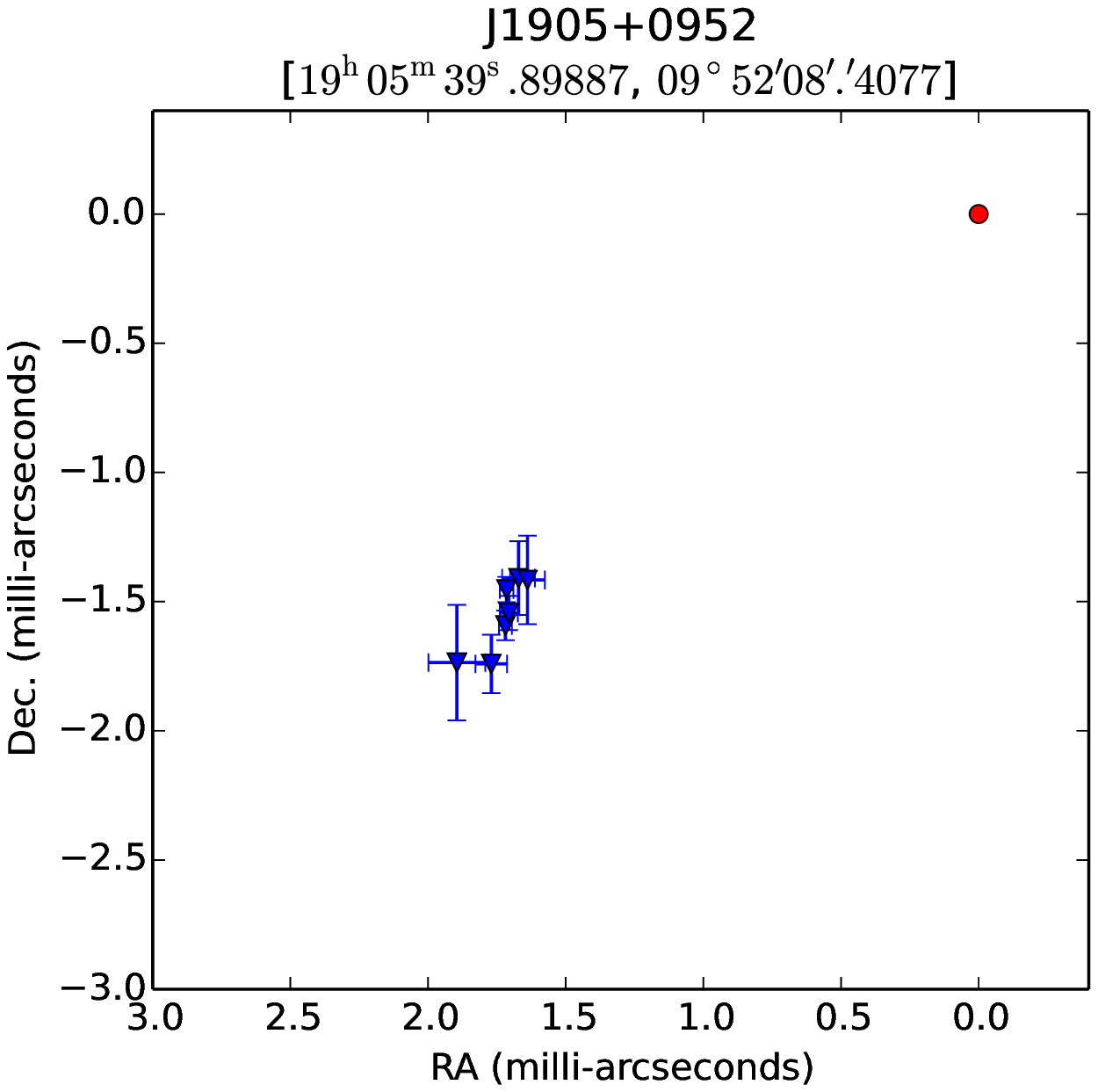} 
\caption{The measured offsets of the check sources from their known positions, as measured by the EVN and VLBA. The top panel shows the difference between the position used at correlation for J1907+0907 (blue), and the observed position (red) when phase-referenced to J1905+0952 in EVN experiments. The bottom panel shows the difference between the position used at correlation for J1905+0952 (red) and the observed position (blue) when phase referenced to J1907+0907 in VLBA experiments.  We use the mean offsets to correct the EVN measured positions to the VLBA frame.}
\label{fig:check_source}
\end{figure}

While all astrometric positions were measured relative to a phase calibrator, the VLBA and EVN used different phase reference sources (J1907+0907 and J1905+0952, respectively). They were also observed at different frequencies, so any core shift in the calibrator due to a resolved jet would result in an astrometric shift in the target image.  We therefore compared the relative positions of the two calibrator sources in each data set, which was possible since in each case the phase reference calibrator for one array was used as the check source for the other. In Figure~\ref{fig:check_source} we show the relative positions of each calibrator as measured by the EVN (in red) and VLBA (in blue). We determined a mean offset of the EVN frame relative to that of the VLBA of $-2.14 \pm 0.14$\,mas in R.A.\ and $1.87 \pm 0.13$\,mas in Dec., and a mean offset of the VLBA frame relative to that of the EVN of $1.73 \pm0.08$\,mas in R.A.\ and $-1.56 \pm0.13$\,mas in Dec..

\begin{figure}
\centering 
\includegraphics[width=7cm,angle=0,trim = 0mm 25mm 0mm 11mm,clip]{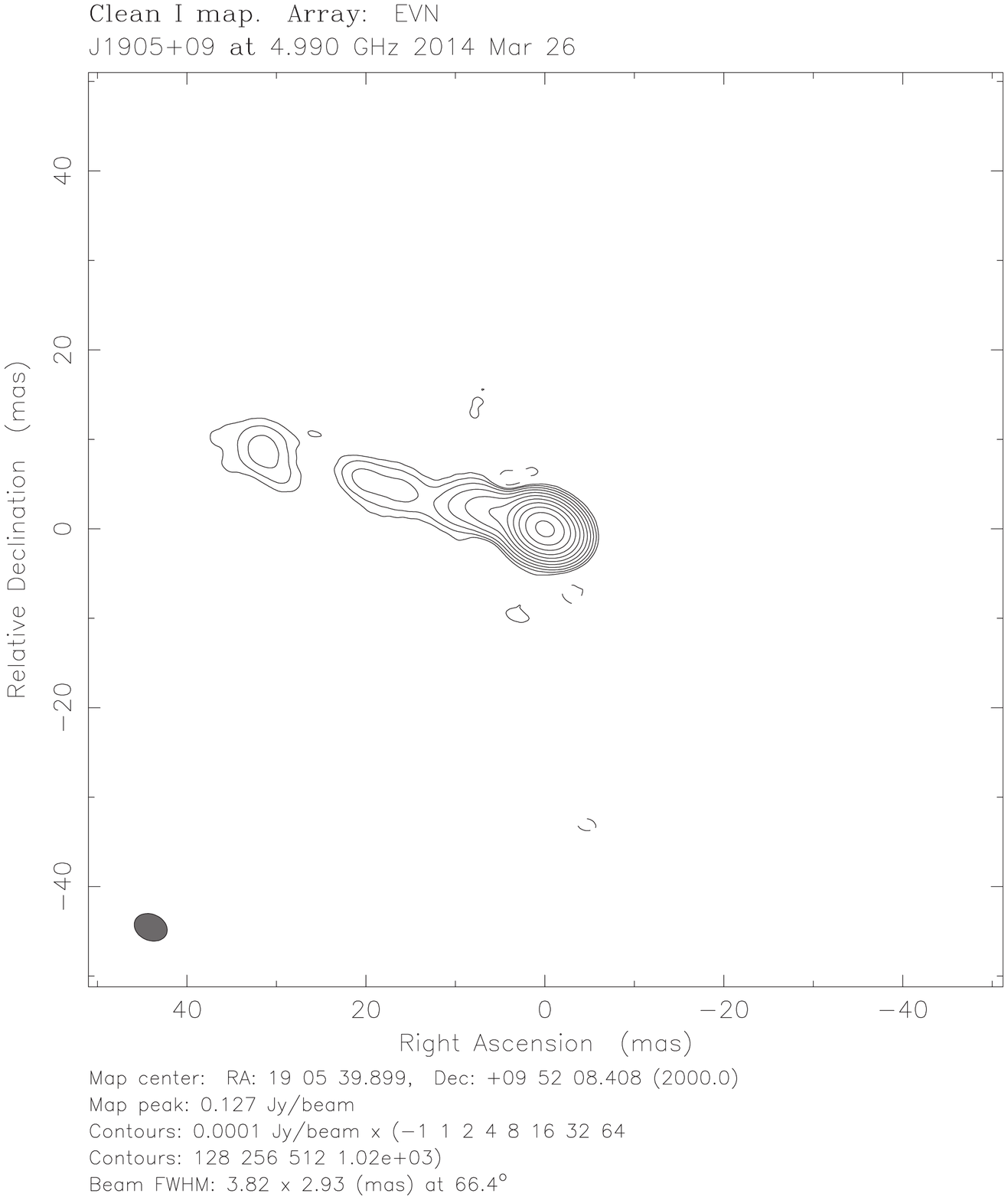} 
\caption{5\,GHz EVN image showing the resolved jet of the
  calibrator source J1905+0952. The peak brightness is 127\,mJy\,bm$^{-1}$, and contours are at levels of $-1,1,2,4,8,16,32, ..., 512$ times the lowest contour of 0.1~mJy~bm$^{-1}$.  The offsets measured in Fig.~\ref{fig:check_source} are not aligned with the jet axis, but could potentially be explained by a combination of the resolved calibrator and unmodelled tropospheric delays that become significant at low elevations.}
\label{fig:J1905}
\end{figure}

The nominal uncertainties on the known calibrator positions\footnote{\url{http://astrogeo.org/}} are $<0.3$\,mas, so cannot alone explain the measured offsets.  However, J1905+0952 was resolved by the EVN (Figure~\ref{fig:J1905}), suggesting that the core location measured at 5\,GHz by the EVN could be shifted downstream along the jet axis as compared to the 8.4-GHz position measured by the VLBA.  The poorly-modelled tropospheric delay in the EVN observations could have introduced a positional shift in declination, and the combination of these two effects, in addition to the high winds prevailing at many of the EVN stations during the observations, could potentially explain the observed shift between the EVN and VLBA frames.

To enable a valid comparison of the positions measured by the two arrays, we therefore attempted to align the two sets of observations onto the same astrometric frame.  Since the offsets measured independently by each telescope overlapped within the uncertainties, then through the rest of this paper we use the average offsets of $\Delta \alpha=1.9$ mas and $\Delta \delta= -1.7$ mas to correct the measured EVN positions to the VLBA reference frame.

\section{Results and Analysis}
\label{section:analysis}

\begin{figure*}
  \centering
 \vspace{30mm}
\includegraphics[width=18cm,angle=0]{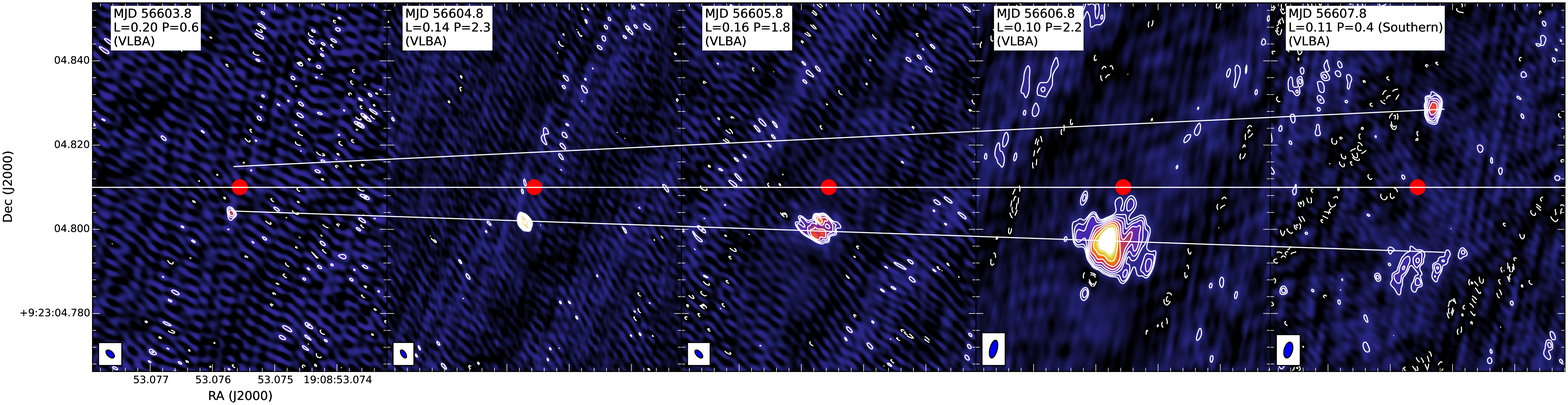} 
\includegraphics[width=09cm,angle=0]{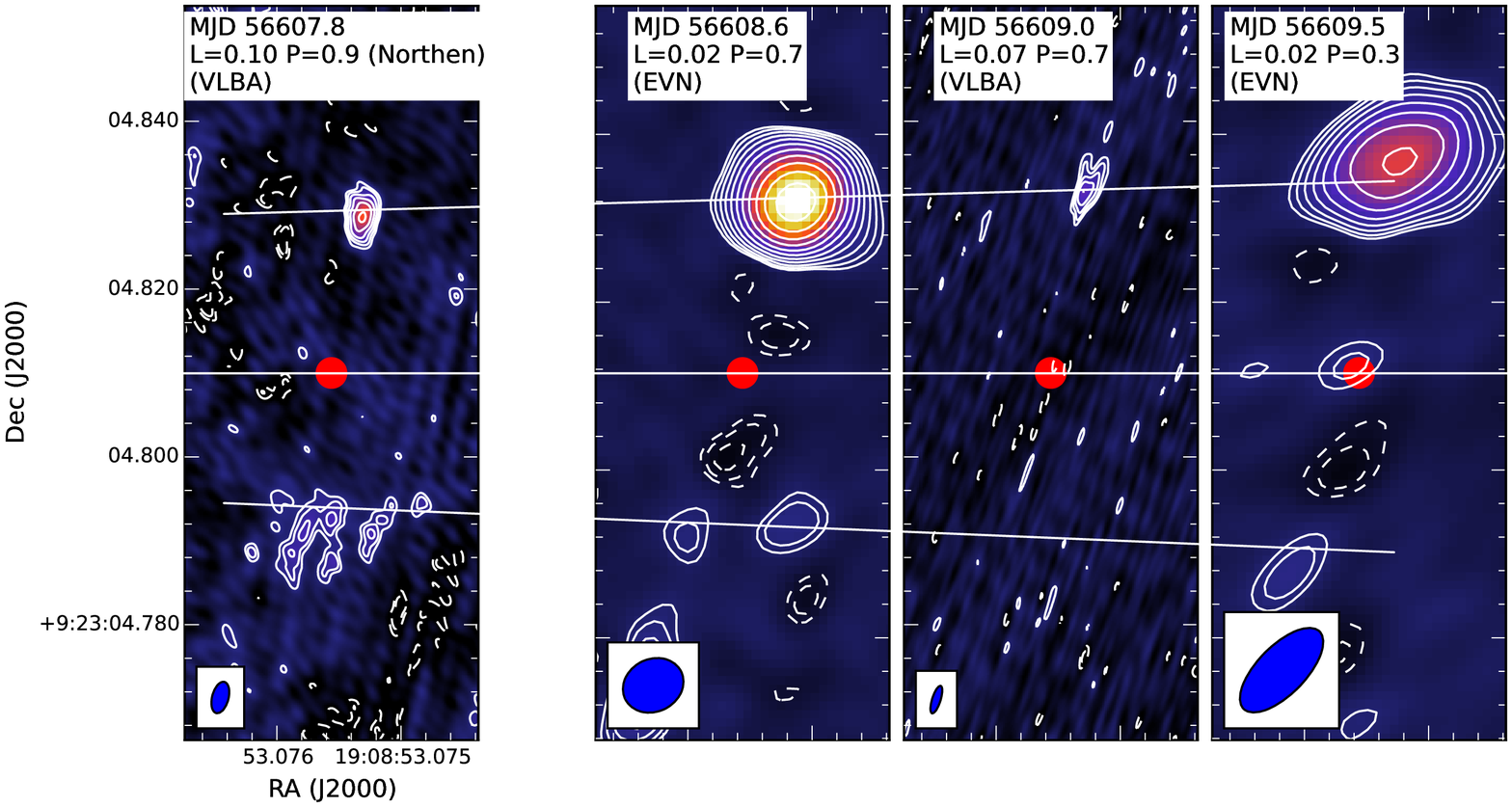} 
  \caption{VLBI maps of \xtenineteen\ during the 2013 outburst, all taken following the hard-to-soft X-ray state transition. The top panels highlight the evolution of the southern component and span 4 days in time, up to its disappearance following the VLBA epoch on MJD 56607.8. The bottom panels highlight the evolution of the northern component from its first appearance on the same day (MJD 56607.8), and span 1.7 days in time. L and P indicate the lowest and peak contours in units of mJy bm$^{-1}$, with contour levels of $\pm(\sqrt{2})^{n}$, where n=3,4,5,6,.... The red circle indicates the estimated position of the binary (Section \ref{sec:propermotions}) and the white lines show the fitted motions of the respective components.  We see clear expansion of the southern component over time, and a delay in the appearance of the northern component.}
\label{fig:images}
 \end{figure*}

The first VLBI observations were taken approximately eight days after the first radio detection of the 2013 outburst (the images are shown in Figure~\ref{fig:images}). The high-resolution observations initially detected a relatively compact, slightly resolved radio component, with a fitted angular size of $2.2\pm0.8$\,mas. Over the course of the next four days the component expanded and moved towards the south. As it did so, the component developed a more complex substructure. Rather than a single, resolved Gaussian, a ``core-halo'' morphology was detected on MJD 56606.8, with a broad, diffuse structure surrounding an inner hot-spot. Four days after the initial detection, the southern component had expanded so much that its surface brightness began to fade below detectability.

As the southern component expanded and became resolved out, a new component appeared $\sim20$~mas to the north.  This northern component moved in the opposite direction, and also started to expand. However, it did not appear to show quite the same level of complexity as the southern component; although two of the VLBI observations were taken at a lower resolution with the EVN (due to the lower observing frequency and shorter baselines), the two VLBA epochs showed a much more compact morphology, without the development of a prominent core-halo structure as seen in the southern component. Furthermore, both the peak and integrated flux of this northern component remained much lower than those of its southern counterpart.

The rise and fall of the large radio flare reported by \citet{2015MNRAS.451.3975C} are contemporaneous with our VLBI observations, and can therefore be explained not as a single event, but as the brightening and fading of two distinct, resolved ejecta. The two swings in the electric vector position angle (EVPA) of the linear polarization detected by \citet{2015MNRAS.451.3975C} could also be related to the appearance and evolution of the different components.  This will be explored in more detail in Section~\ref{sec:comparison}.

\subsection{Motion of resolved components}
\label{sec:propermotions}

\begin{table}
\footnotesize
\caption{Astrometric results}
\begin{center}
\addtolength{\tabcolsep}{-3pt}
\begin{tabular}{lcc}
\hline
Epoch 	&	 \multicolumn{2}{c}{Peak position}	\\
(MJD) 	&	 RA (19h 08m ss.sssssss) 	&	Dec (09\degree 23' ss.ssssss)	\\
\hline					
56603.78	&	 $53.0756925 \pm 0.00018s$	&	$04.804002 \pm 0.0027"$	\\
56604.62	&	N/A 	&	 N/A 	\\
56604.82	&	 $53.0757174 \pm 0.00014s$	&	$04.801382 \pm 0.0022"$	\\
56605.79	&	 $53.0757372 \pm 0.00017s$	&	$04.799978 \pm 0.0026"$	\\
56606.80	&	 $53.0757841 \pm 0.00017s$	&	$04.797786 \pm 0.0026"$	\\
56607.78$^{S}$	&	 $53.0757905 \pm 0.00020s$	&	$04.791226 \pm 0.0030"$	\\
56607.78$^{N}$	&	 $53.0753187 \pm 0.00020s$	&	$04.828601 \pm 0.0030"$	\\
56608.64	&	 $53.0751346 \pm 0.00020s$	&	$04.832090 \pm 0.0030"$	\\
56609.00	&	 $53.0752967 \pm 0.00015s$	&	$04.831471 \pm 0.0023"$	\\
56609.54	&	 $53.0752819 \pm 0.00062s$	&	$04.836453 \pm 0.0093"$	\\
56610.90	&	N/A 	&	N/A	\\
56612.92	&	 N/A 	&	N/A	\\
56617.90	&	 N/A 	&	N/A	\\
56742.25	&	 N/A 	&	N/A	\\
\hline
\end{tabular}
\end{center}
\label{table:astrometry}
\end{table}

\begin{figure}
\centering 
\includegraphics[width=9cm, clip=true, trim= 0cm 0cm 0cm 0cm]{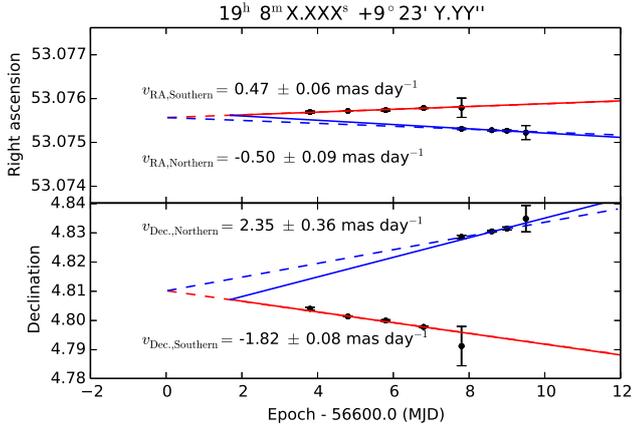}
\caption{Astrometric positions of the components. The red and blue
  lines are weighted best fits to the southern and northern
  components respectively, assuming they are ejected simultaneously
  and exhibit ballistic motion. The dashed lines indicate the best fit with
  $t_0$ as a free parameter (fitted as MJD 56600.0), and correspond to the proper motions stated on the plot.  The solid lines used a fixed
  ejection time (corresponding to the prompt radio flare at MJD
  56601.7), and give a slightly higher velocity in declination for the
  northern component of
  $v_{\rm{Northern}}=3.35\pm0.45$~mas~day$^{-1}$.  Both components appear to exhibit ballistic motion.}
\label{fig:positions}
\end{figure}

Our VLBI observations were triggered too late to detect the ``steady" jet emission from the core and we can place an upper limit of $\lesssim30~\mu Jy$ on the core brightness in our deepest epoch on MJD 56608--9 (EVN observation RR007B). While this precludes an unambiguous determination of the location of the central binary system, we are nonetheless able to estimate the core position by extrapolating the motions of the ballistically-moving ejecta backwards in time until they intersect. To do so, we used the measured positions of the peak flux density in each image to estimate the precise position of the binary at each epoch (Table~\ref{table:astrometry}), and then independently fit the positions of the northern and southern components with a linear motion (Figure ~\ref{fig:positions}).  While it is possible that the northern and southern components were initially ejected at different epochs, this seems unlikely based on our knowledge of typical X-ray binary jet properties. It is also unlikely that the components were ejected before the X-ray state change, which began on or prior to MJD 56597.92 \citep{Zhang15}. 

The fitted proper motion of the southern component was $\mu_{\rm{S}}=1.88\pm0.08$ mas day$^{-1}$ at a position angle of $167.5\pm0.6^{\circ}$ east of north. Our fit to the motion of the northern component gave $\mu_{\rm{N}}=2.40 \pm 0.35$~mas~day$^{-1}$ along a position angle of $-11.1\pm2.8^{\circ}$, consistent with it being the bipolar counterpart of the southern component. This led to an inferred position for the central binary of (J2000) RA~$19^{\rm{h}}08^{\rm{m}}53\fs 07556\pm 0\fs 00004$,
Dec. $+9^{\circ}23'04\farcs{810}\pm0\farcs{002}$ (marked as a red circle in Figure~\ref{fig:images}) and an ejection date of $t_0=~\rm{MJD}~56600.0\pm0.7$. We note that the southern component first appears significantly closer to the estimated binary position than the first detection of its northern counterpart.

Assuming an intrinsically symmetric ejection event we can use the approaching and receding proper motions ($\mu_{\rm a}$ and $\mu_{\rm r}$, respectively) to constrain the intrinsic jet velocity and inclination angle via
\begin{equation}
    \mu_{\rm r,a} = \frac{\beta \sin \theta}{1\pm\beta\cos\theta},
\end{equation}  
which leads to the constraint
\begin{equation}
    \beta\cos\theta = \frac{\mu_{\rm a}-\mu_{\rm r}}{\mu_{\rm a}+\mu_{\rm r}},
\end{equation}
where $\beta=v/c$ is the jet speed as a fraction of the speed of light and $\theta$ is the inclination angle of the jet axis to the line of sight. From the measured proper motions and assuming that the faster-moving northern component is the approaching one, this would imply $\beta\cos\theta=0.12\pm0.07$.  Therefore, since both $\beta<1$ and $\cos\theta<1$ we constrain the intrinsic velocity to be $<0.19c$ and the inclination angle $\theta>79$\degree, (both at $1\sigma$ confidence). Moreover, at a distance of 8 kpc, the projected separation speed of the components in the plane of the sky is $<0.2c$, reinforcing the conclusion of a low jet velocity if the components are intrinsically symmetric ejecta modified by relativistic boosting and light travel time effects.

Given that our derived ejection date was several days before any significant changes in the overall radio light curve, we also re-fit the component motions assuming the ejection to have taken place at the time of the rapid radio flare seen by AMI-LA on MJD\,56601.7 \citep{2015MNRAS.451.3975C}. Given the high-cadence VLBA sampling of the southern component, this assumption only significantly changed the velocity of the northern knot, which increased to 3.4\,mas\,day$^{-1}$ (see Figure~\ref{fig:positions}).  Such a high proper motion relative to the southern component would suggest that if the ejecta did correspond to the initial radio flare at MJD\,56601.7, then the northern component is likely to be the approaching one.

\subsection{Expansion of resolved components}

\begin{figure}
\centering 
\includegraphics[width=8.5cm]{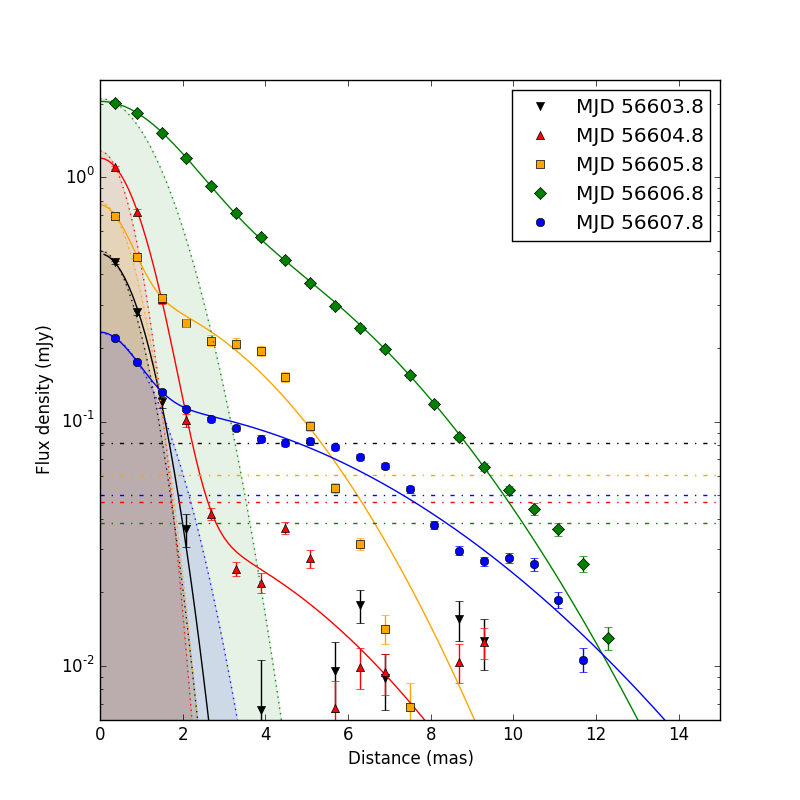} 
\includegraphics[width=8.5cm]{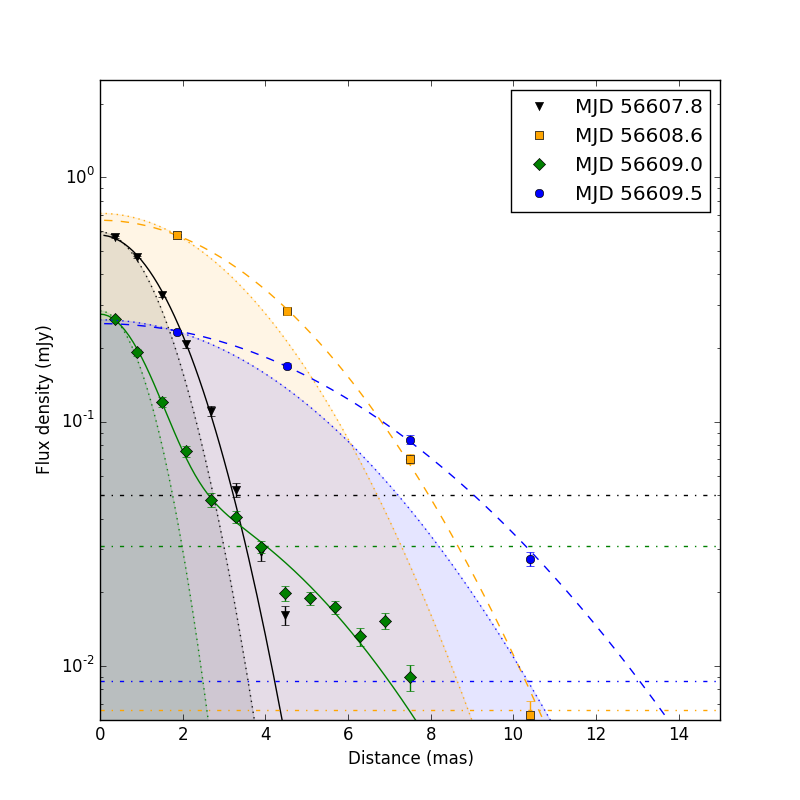} 
\caption{Radial profiles of the extended southern ({\it top panel}) and northern ({\it bottom panel}) ejecta, derived using the {\sc AIPS} task {\sc iring}.  We fit each profile with one-dimensional Gaussian models centred on the position of peak flux in each image. The solid lines are fits to the 8.4-GHz VLBA data and the dashed lines are fits to the 5-GHz EVN data. The horizontal dot-dashed lines show the RMS noise level of each epoch. The shaded regions represent the minimum beam size in each image, as a guide to the intrinsic angular resolution of the images. In several of the epochs we see clear evidence for two components, corresponding to a core-halo morphology.}
\label{fig:iring}
\end{figure}

As discussed in Section~\ref{section:analysis}, the two components expanded as they moved outwards. Figure \ref{fig:images} shows that the southern knot was initially very compact. It then expanded as it moved away from the central binary system, developing a bright core and extended halo before becoming resolved out by MJD 56608. The initial VLBA detection of the northern component was slightly resolved, and by the time of the second observation it had expanded to $8.3\times 1.1$\,mas$^2$ in size (Table~\ref{table:sizes}). Unfortunately, the lower resolution of the EVN makes it difficult to further quantify the expansion of the northern component.

\begin{table*}																	
 \caption{\textsc{jmfit} results showing the fitted sizes of the Southern \& Northern components.  Each component was fitted with either one or two Gaussians, as indicated.  Two-Gaussian fits correspond to the epochs with a core-halo morphology. $\theta_{\rm maj}$, $\theta_{\rm min}$ and $PA$ are respectively the major and minor axes and position angle of the fitted elliptical Gaussians.}
 \label{table:sizes}
\begin{center}																	
 \begin{tabular}{llcccccccc}																	
 \hline																	
& Epoch	&	$\theta_{\rm maj,1}$	&	$\theta_{\rm min,1}$	&	$PA_1$	&	Total flux$_1$	&	$\theta_{\rm maj,2}$	&	$\theta_{\rm min,2}$	&	$PA_2$	&	Total flux$_2$	\\
& (MJD)	&	(mas)	&	(mas)	&	(deg.)	&	(mJy)	&	(mas)	&	(mas)	&	(deg.)	&	(mJy)	\\
 \hline																	
 \parbox[t]{2mm}{\multirow{5}{*}{\rotatebox[origin=c]{90}{Southern}}} & $56603.78\pm0.02$	&	$\phn3.79\pm0.40$	&	$\phn2.10\pm0.22$	&	$158\pm\phn7$	&	$\phn0.74\pm  0.13$	&		&		&		&		\\
&$56604.82\pm0.05$	&	$\phn6.91\pm2.84$	&	$\phn2.52\pm1.04$	&	$163\pm14$	&	$\phn0.89\pm0.41$	&	$2.55\pm0.09$	&	$1.94\pm0.07$	&	$\phn30\pm\phn5$	&	$3.10\pm0.15$	\\
&$56605.79\pm0.03$	&	$\phn7.98\pm0.65$	&	$\phn6.21\pm0.51$	&	$161\pm13$	&	$\phn9.08\pm0.78$	&	$2.05\pm0.56$	&	$1.02\pm0.28$	&	$\phn46\pm15$	&	$<0.27$	\\
&$56606.80\pm0.06$	&	$14.24\pm0.92$	&	$\phn9.37\pm0.60$	&	$103\pm\phn6$	&	$10.94\pm0.74$	&	$4.22\pm0.11$	&	$3.48\pm0.09$	&	$160\pm\phn5$	&	$4.69\pm0.16$	\\
&$56607.78\pm0.04$	&	$15.11\pm2.18$	&	$11.79\pm1.70$	&	$160\pm\phn5$	&	$\phn4.72\pm0.71$	&		&		&		&		\\
 \hline																	
  \parbox[t]{2mm}{\multirow{4}{*}{\rotatebox[origin=c]{90}{Northern}}} & $56607.78\pm0.04$	&	$\phn4.69\pm0.41$	&	$\phn2.65\pm0.23$	&	$175\pm\phn6$	&	$\phn0.97\pm0.13$	&		&		&		&		\\
&$56608.64\pm0.17$	&	$\phn8.32\pm0.07$	&	$\phn7.72\pm0.10$	&	$112\pm\phn5$	&	$\phn0.94\pm0.13$	&		&		&		&		\\
&$56609.00\pm0.05$	&	$\phn8.30\pm2.33$	&	$\phn1.14\pm0.32$	&	$149\pm\phn4$	&	$\phn0.30\pm0.11$	&	$5.83\pm0.82$	&	$1.94\pm0.27$	&	$174\pm\phn3$	&	$0.70\pm0.13$	\\
&$56609.54\pm0.06$	&	$14.22\pm0.44$	&	$\phn9.33\pm0.29$	&	$\phn14\pm\phn3$	&	$\phn0.48\pm0.02$	&		&		&		&		\\
 \hline																	
 \end{tabular}																	
 \end{center}																	
 \end{table*}													
The emitting regions became resolved over the course of the observing campaign, and were not always well approximated by standard elliptical Gaussian models, making it difficult to quantify the source size and hence expansion rate.  To get an initial idea of the extension of the components at each epoch, we first used the \textsc{AIPS} task \textsc{iring}.  This task integrates the flux contained within successive concentric annuli, which we centred on the position of peak flux density in a given image.  This provided an azimuthally-averaged estimate of the radial extent of each component, providing a first-order size estimate, which was particularly useful for the most heavily-resolved epochs.  We chose the width of each annulus to be 3 pixels (equivalent to 0.6--0.9\,mas for the VLBA and 3\,mas for the EVN), and the results are shown in
Figure~\ref{fig:iring}. In an attempt to parametrize the resulting radial profiles, we tried fitting them with a variety of simple one-dimensional models (Gaussian, Lorentzian and power law functions) and examined the residuals. We found the best fits when using either a one- or two-component Gaussian
function centred on the peak emission, which gave residuals consistent with the image noise.

The two-Gaussian fits correspond to a `core-halo'
morphology, suggesting that while some of the flux remained
unresolved it was surrounded by more diffuse emission. This `halo'
seemed to extend up to $\sim20$~mas in diameter and was not
present in the first epoch. More detailed inspection of the individual images (Fig.~\ref{fig:images}) showed the halo to be relatively isotropic (i.e.\ not significantly elongated parallel or perpendicular to the position angle of the jet) and to move with the bulk motion of the ejecta. 																			

\begin{figure}
\centering 
\includegraphics[width=8.5cm]{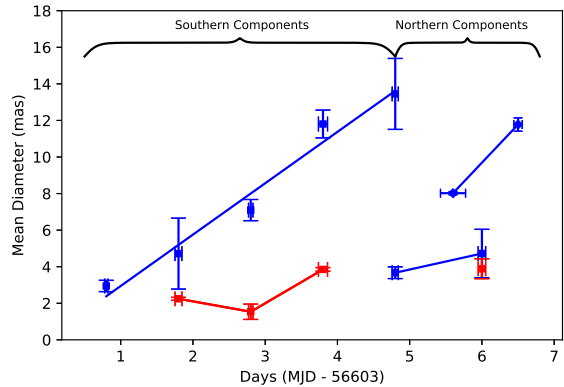} 
\caption{Expansion of the two ejected components. The blue points are the mean diameter of the surrounding `halo' structure and the red points are the epochs that also displayed an inner `core' component. The southern halo appears to expand at a constant rate.} 
\label{fig:expansion_rate}
\end{figure}

Using the \textsc{aips} task \textsc{jmfit} we fitted the images with either one or two elliptical Gaussians depending on the best \textsc{iring} fit (as listed in table~\ref{table:sizes}). The results of this two-dimensional image-plane analysis showed a slight tendency for the major axis of the elliptical Gaussians to be aligned along a position angle close to the jet direction of $\sim 167$\degree\ E of N. When fitting the mean sizes of each component over time, the outer halo structure of the southern component was found to expand linearly at a rate of
$3.2\pm0.5$~mas~day$^{-1}$ (Fig.~\ref{fig:expansion_rate}). Midway through the expansion, the core component rebrightened and became marginally resolved, disappearing again $\sim36$~hrs later. For the northern component we only fit
an expansion rate to the 5-GHz observations, as it is not meaningful to compare sizes at different wavelengths.  This gave an expansion rate of $2.9\pm0.1$~mas~day$^{-1}$.  Thus, the northern and southern components appeared to have similar expansion rates of $\sim 3$~mas~day$^{-1}$, corresponding to roughly $0.15c$ at 8\,kpc.
 
In the VLBA epoch on MJD 56609.0, we also attempted a two-Gaussian fit to the northern component, guided by the {\sc iring} results (see Fig.~\ref{fig:iring}), which tentatively suggested a core-halo structure. 
To test the reality of this hypothesis, we also conducted model fitting of the visibility data within the \textsc{Difmap} software package \citep{Shepherd97}. While \textsc{difmap}
does not return a formal uncertainty on the derived parameters, by model-fitting in the visibility domain it can rule out any spurious structures arising from imaging artefacts. In this case the \textsc{Difmap} fits did warrant a two-Gaussian fit, giving us more confidence in the development of a similar diffuse ``halo" structure around the northern component.

\section{Discussion}
\label{section:discussion}

Our VLBI observations were made following the hard-to-soft X-ray spectral state transition at the peak of the 2013 outburst of \xtenineteen.  We detected and tracked a set of ballistically-moving ejecta that expanded on moving away from the core, and directly resolved the jets perpendicular to the direction of motion.  Although our VLBI observations did not detect the ``steady" jet emission from the core that is typically seen in the hard/plateau state \citep[e.g. \grs;][]{2000ApJ...543..373D,2010MNRAS.401.2611R}, we were able to infer its location by extrapolating the proper motions of the ballistically-moving components back to zero separation.

To explain the observed properties of the evolving ejecta, we now briefly review our main findings in the context of the standard models for X-ray binary jets.  We then combine our imaging results with the radio light curves, spectra and polarimetry presented by \citet{2015MNRAS.451.3975C}, to arrive at a preferred scenario for the observed evolution of the jets in \xtenineteen.

\subsection{Key observational signatures}

\subsubsection{North/south asymmetry}

There are clear differences in the properties and evolution of the northern and southern components.  The southern component appears four days earlier than its northern counterpart, and at its maximum flux density is also significantly brighter than the peak emission from the northern component.  Furthermore, the southern component persists for longer, and appears to be more extended.  

Differences in the flux densities of approaching and receding ejecta in X-ray binary jets are often ascribed to Doppler boosting, such that the approaching component appears significantly brighter than the receding component.  This explanation would lead us to infer that the southern component was approaching and the northern component receding.  While this would initially seem to explain both the observed differences in peak flux density and the delayed appearance of the northern component (due to light travel time effects), such an explanation would be at odds with the lower proper motion measured for the southern component (Section \ref{sec:propermotions}).

Assuming that the two components were intrinsically identical, were ejected at the same time, and moved ballistically, the observed proper motions of both components then constrain $\beta\cos\theta=0.12\pm0.07$.  This implies a low jet speed, a motion close to the plane of the sky, or both.  We note that the X-ray spectral fitting of \citet{Miller09}, \citet{Tao15}, and \citet{Zhang15} determined an inclination angle of the inner disc (presumably aligned with the black hole spin and hence with the jet axis) that lies in the range 20--53\degree, making it unlikely that the jets are close to the plane of the sky.  At a fiducial distance of 8\,kpc, a proper motion of 1\,mas\,d$^{-1}$ would imply a projected velocity of $0.046c$ in the plane of the sky.  Thus, unless \xtenineteen\ is significantly more distant than 8\,kpc, the measured proper motions of $\mu_{\rm N}=2.40\pm0.35$\,mas\,d$^{-1}$ and $\mu_{\rm S}=1.88\pm0.08$\,mas\,d$^{-1}$ indeed imply a jet speed significantly lower than observed in other BH XRBs such as \grs\ \citep{1994Natur.371...46M}, \gro\
\citep{1995Natur.375..464H}, or H1743-322 \citep{2012MNRAS.421..468M}.

Furthermore, under the above assumptions and using the measured constraint on $\beta\cos\theta$, we can determine the expected flux density ratio between the northern and southern ejecta. At equal angular separations \citep[e.g.][]{2004ApJ...603L..21M}, the ratio of flux densities should be given by
\begin{equation}
\frac{S_{\rm app}}{S_{\rm rec}} = \left(\frac{1+\beta\cos\theta}{1-\beta\cos\theta}\right)^{k-\alpha},
\end{equation}
where $k=3$ for discrete ejecta and $k=2$ for a continuous, steady jet, and $\alpha$ is the radio spectral index.
Assuming discrete components and an appropriate spectral index of $-0.6$, we would then expect a flux density ratio between the northern and southern components of 1.3--4.0.  Instead, the slightly faster-moving northern component (see Fig.~\ref{fig:positions}) has a flux density that is significantly lower than that of the southern component.  Specifically, on MJD 56607.8, when both components have the same angular separation from the core within uncertainties, the integrated flux from the southern component is a factor $1.9\pm0.3$ greater than that of the northern component.  In summary, the standard special relativistic effects applied to explain the behaviour of the jets in sources such as \grs\ do not appear to be solely responsible for the observed behaviour of \xtenineteen.

The differences in the observed expansion of the two components also argue that we are not simply seeing two intrinsically symmetric jets whose observed evolution is dictated by standard relativistic boosting and light travel time delays.  While our VLBA imaging shows that the southern component expands to a size of 13\,mas by MJD 56607.8 (after which time it becomes resolved out), the northern component reaches a size of only 6\,mas before it fades below detectability.  The EVN observations, taken at a lower frequency and with shorter baselines, and hence sampling larger spatial scales, do show the northern knot to be slightly more extended.  However, in the absence of similar EVN data on the southern component, we have no valid point of comparison at 5\,GHz.

\subsubsection{Comparison to photometry and polarimetry}
\label{sec:comparison}

\begin{figure*}
\centering 
\includegraphics[width=\textwidth]{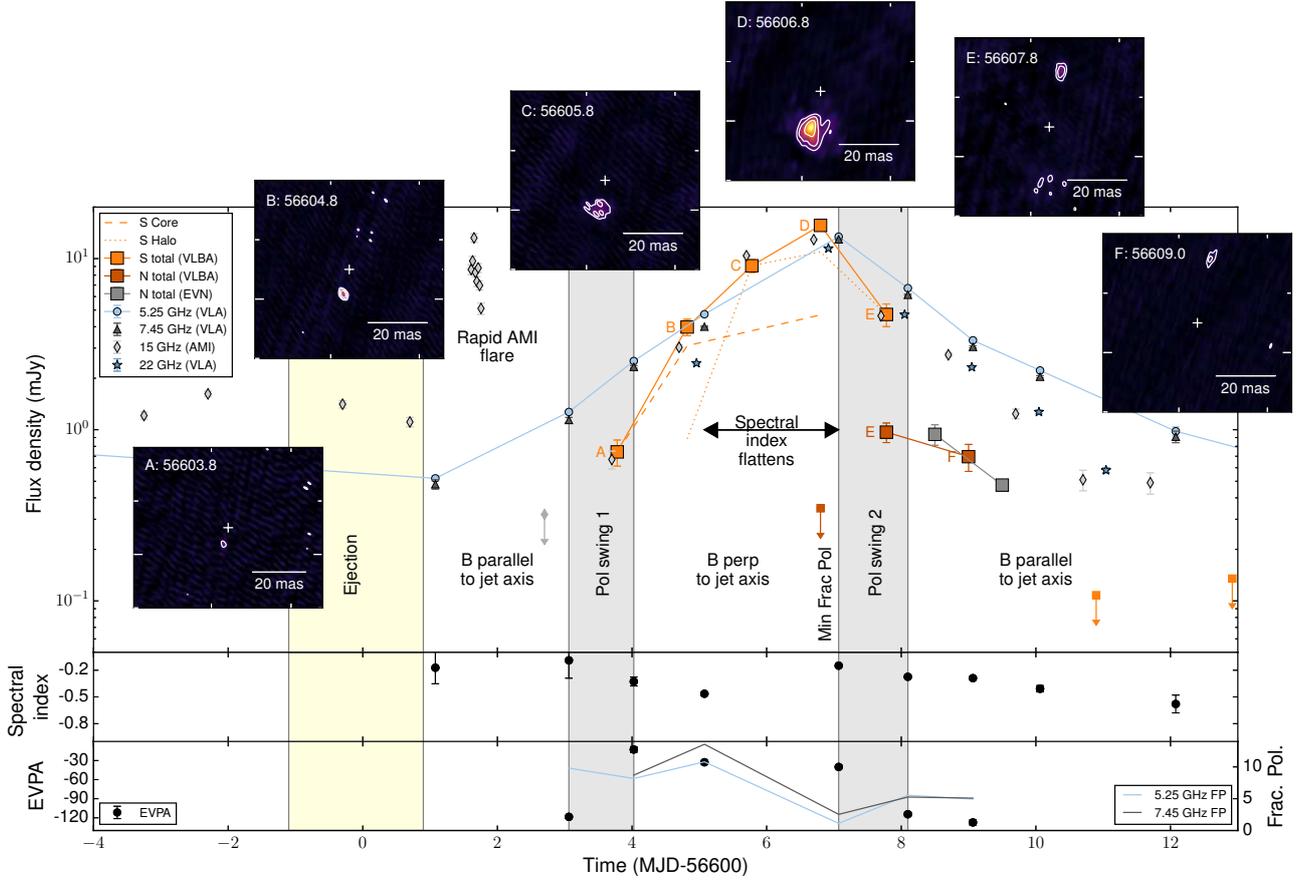} 
\caption{The evolution of the 2013 outburst of XTE J1908+094.  {\it Main panel} shows the overall radio light curves, with the total flux measured by the VLBA (orange, red squares for S and N components), EVN (grey squares), VLA (5.25\,GHz, 7.45\, GHz and 22\,GHz represented by blue circles, grey triangles and blue stars, respectively) and AMI (15\,GHz; grey diamonds). The range of likely ejection times is shaded yellow, and the periods where 90$^{\circ}$ swings of the EVPA were observed are shaded grey.  {\it Insets} show the VLBA images over the six epochs (A--F) where components were detected, each of which is associated with one of the indicated light curve points on the main panel.  Contours are at levels of $2^n$ times the rms noise, where $n=3,4,5,...$.  {\it Middle panel} shows the evolution of the radio spectral index and {\it lower panel} shows the evolution of the intrinsic EVPA (left axis) and the fractional polarisation (right axis), all taken from \citet{2015MNRAS.451.3975C}.  The large radio flare from MJD 56602--56610 is accompanied by strong evolution of the VLBI-detected components.  The polarization angle swings correspond to the appearance and disappearance of the southern component, and the flattening spectral index corresponds to the re-energisation of the core of the southern component.}
\label{fig:lcs_all}
\end{figure*}

While our VLBI imaging data provide important information as to the behaviour of the jets during the outburst, we can obtain further context from the photometric, spectral and polarimetric observations of \citet{2015MNRAS.451.3975C}.  The key points in the source evolution are shown in Fig.~\ref{fig:lcs_all}.  The outburst began with fairly steady radio emission at around 1--2\,mJy\,\perbeam.  This persisted until MJD 56601, when AMI-LA detected a short-lived 15-mJy flare.  The radio emission subsequently decreased to 0.3\,mJy\,\perbeam, before rising again from MJD 56603 to 56607.  It then decayed away over the course of the next week, until the source was no longer detected.

The rise in flux density from MJD 56603--56607 coincides with the VLBI detections of the southern component, which gets brighter and expands over time, reaching its peak integrated flux on MJD 56606.8, before it fades and becomes resolved out (Fig.~\ref{fig:images}).  Prior to the peak of the flare, our VLBI images recover virtually all of the radio emission detected by AMI and the VLA.  Our first VLBA epoch recovers all of the AMI flux, although misses a few tens of percent of the flux seen several hours before and after by the VLA.  During the decay phase of the flare, following the peak on MJD 56606.8, the southern component is resolved out by the VLBA, although from the light curves in Fig~\ref{fig:lcs_all}, we infer that it continues to dominate the integrated radio emission.

During the rise and decay of the main flare, \citet{2015MNRAS.451.3975C} detected significant swings in the intrinsic electric vector position angle (EVPA) of the linear polarization.  This first changed during the rise phase, from $-118$\degree\ on MJD 56603.1 to $-13\pm5$\degree\ on MJD 56604.0.  Over the next few days, as the southern component evolved and expanded, the intrinsic EVPA gradually evolved to $-40\pm3$\degree\ by MJD 56607.1, after which it showed a second large swing, to $-114\pm3$\degree\ on MJD 56608.1.  This coincides with the disappearance of the southern component and the appearance of the northern component.  For an optically thin synchrotron source, the EVPA should be perpendicular to the projection of the magnetic field on the plane of the sky.  Since the EVPA was approximately aligned with the observed jet axis (see Section~\ref{sec:propermotions}) on MJD 56604.0, we infer that the magnetic field was perpendicular to the jet axis.  This would be consistent with shock compression, either from internal shocks within the jet, or from external shocks where the jets ran into the surrounding ISM.  The subsequent slow rotation of the EVPA up to MJD 56607.1 corresponds to the expansion of the southern component and the decrease in the fractional polarization.  This suggests that the magnetic fields in the southern knot were becoming less well aligned as the knot expanded.  The second large EVPA swing corresponding to the appearance of the northern component suggests that the magnetic field in the northern component was aligned relatively well with the jet axis.  This cannot be explained as shock compression perpendicular to the jet axis, and instead suggests a more complicated magnetic field geometry in the northern component \citep[see, e.g.,][and discussions therein]{2014MNRAS.437.3265C,2015MNRAS.451.3975C}.

In addition to the photometric and polarimetric results, \citet{2015MNRAS.451.3975C} also detected a steepening spectral index between MJD 56603.1 and 56605.1 as the southern component expanded, but then saw the spectral index flatten again from $-0.5$ to $-0.15$ between MJD 56605.1 and 56607.1, after which it steepened continuously to $\alpha=-0.8$ by MJD 56616.0.  This flattening of the spectral index coincides with the development of the bright hotspot in the southern component on MJD 56606.8 (the ``core'' in the core-halo structure).  This suggests ongoing particle acceleration at the hotspot before the emitting region expands adiabatically and the synchrotron self-absorption turnover moves to lower frequencies, giving rise to the subsequent spectral steepening.  No similar flattening is seen following the appearance of the northern component, although the spectral index seems to stabilise at $\alpha=-0.3$, with the decreasing trend stalling between 56608.1 and 56609.1, just after the northern component first appears on MJD 56607.8.  This may be suggestive of some particle acceleration at the shock giving rise to the northern component.

\subsubsection{Comparison to the X-ray behaviour}
\label{sec:xray}

\begin{figure}
\centering 
\includegraphics[width=\columnwidth]{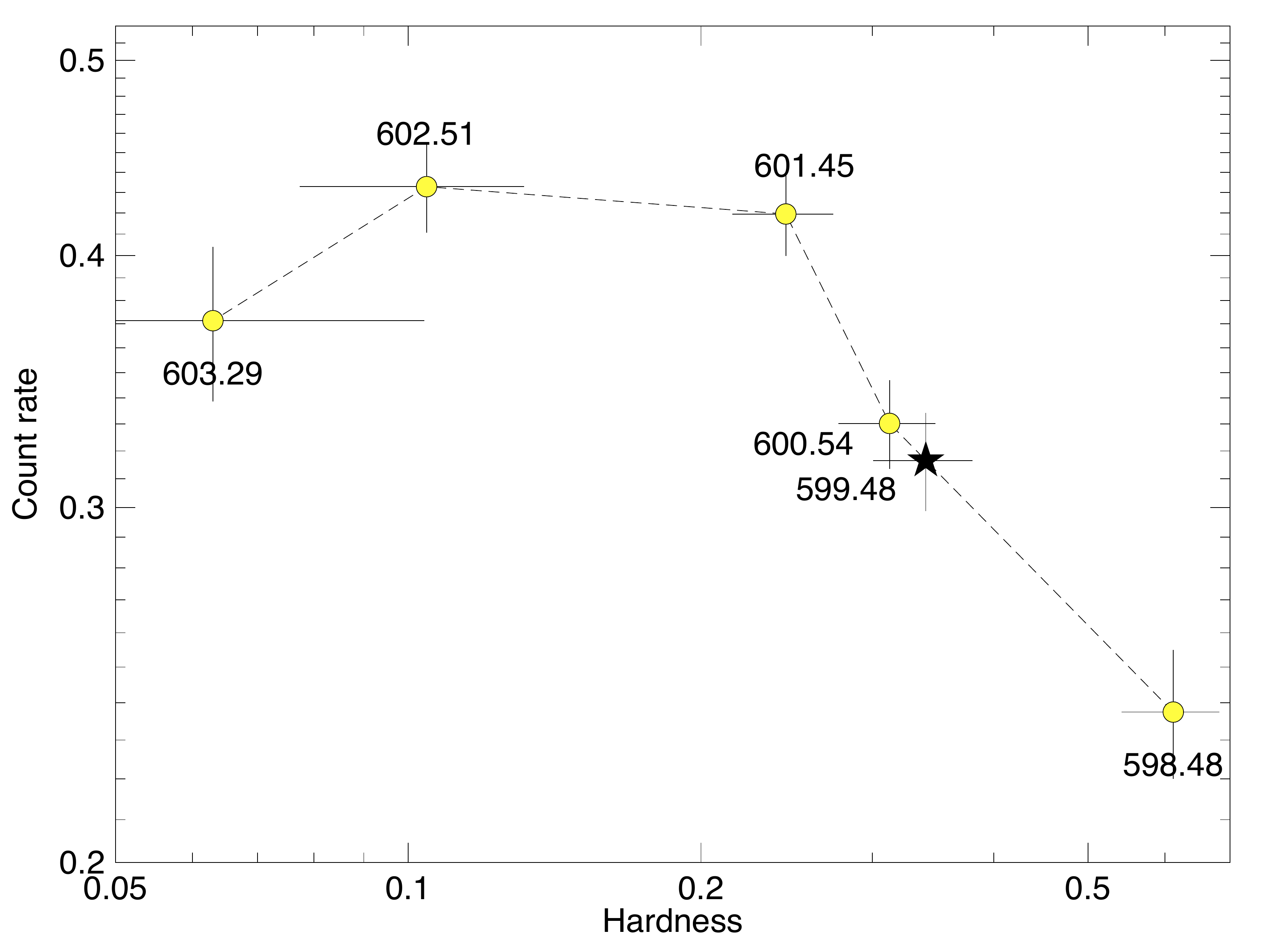} 
\caption{X-ray hardness-intensity diagram made from the publicly-available MAXI data.  Hardness is the ratio of counts in the 6--20 and 2--6\,keV bands, and intensity is the total 2--20\,keV count rate.  The date of each observation is shown as MJD-56000.  The black star shows the observation closest in time to the zero-separation epoch determined in Section~\ref{sec:propermotions}.}
\label{fig:maxi_hid}
\end{figure}

As demonstrated by fig.~4 of \citet{2015MNRAS.451.3975C}, the flare in the VLA and AMI light curves between MJD 56603 and 56620 took place during a period of very soft X-ray photon index, likely corresponding to either the soft-intermediate or soft X-ray spectral state.  Indeed, a subsequent analysis of the X-ray spectral evolution by \citet{Zhang15} suggests that \xtenineteen\ moved from an intermediate state on MJD 56599.5 to a soft state by the time of the following {\it Swift}/XRT observation on MJD 56604.9.  Although the absence of X-ray timing information and the sparsity of the XRT coverage does not allow us to unambiguously determine the exact date of the state transition from the {\it Swift} data alone, the better-sampled MAXI data provide additional constraints.  Plotting a hardness-intensity diagram from the publicly-available MAXI light curves\footnote{\url{http://maxi.riken.jp/mxondem/}} shows a clear softening between MJD 56598.5 and 56603.3 (Fig.~\ref{fig:maxi_hid}).

We find that our inferred zero-separation date for the jet ejecta corresponds to the beginning of the spectral softening, making it plausible that the components did indeed move ballistically between the ejection date and our first VLBI detections.  However, we cannot rule out some deceleration should the ejection have taken place during the latter part of the X-ray spectral softening. The sharp radio flare detected by AMI-LA on MJD 56601 occurs towards the end of the X-ray spectral softening, in the gap between the final intermediate state observation and the first detection of the soft state \citep[as classified from the {\it Swift}/XRT data by][]{Zhang15}.  Given the unknown delay between ejection date and any internal shocks forming and becoming optically thin \citep{Fender09}, we can only infer that the ejection event took place earlier than this.  Finally, with the exception of the first VLBA epoch, all our VLBI observations took place in the soft X-ray spectral state.

The hard X-ray flare detected by NuSTAR around MJD 56605.3 and lasting for 40\,ks \citep{Zhang15,Tao15} occurred during the rise phase of the large radio flare, between our second and third VLBA observations (epochs B and C in Fig.~\ref{fig:lcs_all}).  Although both \citet{Zhang15} and \citet{Tao15} suggested that the flare could have signified the ejection of new transient radio-emitting jet knots, our VLBI imaging shows no new jet component appearing at this time.  The only clear radio changes that could potentially be associated with the hard X-ray flare are the slight flattening of the radio spectral index and the decrease in fractional polarization seen by the VLA between MJD 56605 and 56607, the development of the bright ``core'' in the southern jet component on MJD 56606.8, and the appearance of the northern component on MJD 56607.8.  However, with the available data, we cannot unambiguously tie the hard X-ray flare to any of these events.

\subsubsection{Comparison to the 2002 outburst}
\label{sec:2002}

\begin{figure}
\centering 
\includegraphics[width=0.9\columnwidth,angle=270]{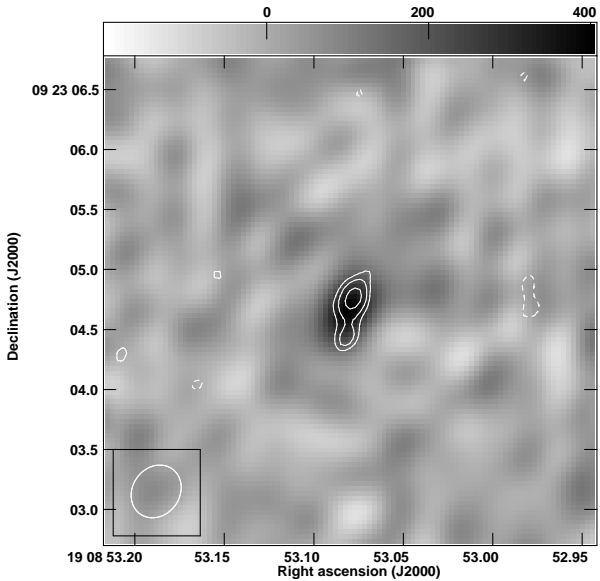} 
\caption{VLA image of \xtenineteen, taken on 2002 April 23.  Contours show the 8.4-GHz data and greyscale shows the 4.9-GHz data.  Contours are at levels of $\pm\sqrt{2}^n$ times the rms noise of 35\,$\mu$Jy\perbeam, with $n=3,4,5,...$.  The 8.4-GHz data shows the source to be resolved into two components separated by $326\pm35$\,mas along a position angle of $-14\pm6$\degree\ E of N.  The overall radio spectrum is steep, with $\alpha=-0.8\pm0.4$. The VLA showed the source to be extended along the same position angle during the 2002 outburst as seen during the 2013 outburst, but with a significantly greater angular separation of the components.}
\label{fig:2002}
\end{figure}

\citet{2002IAUC.7874....1R} detected the radio counterpart of \xtenineteen\ during its 2002 outburst.  The first detection was on March 21, and the source was detected through May 13 \citep{2002IAUC.8029....2R}.  The radio monitoring was relatively sparse, with ten observations over that 53-day period.  This monitoring\footnote{\url{http://www.aoc.nrao.edu/~mrupen/XRT/X1908+094/x1908+094.shtml}} showed that the radio emission peaked on March 30, with non-detections on April 7 and 15, and the source reappearing as a resolved double on April 23 and May 1.  On April 23, the two components had an angular separation of $\sim 0.3$\arcsec\ and a position angle $-14\pm6$\degree\ E of N (Fig.~\ref{fig:2002}).  While this position angle is very similar to that observed in our VLBI data, their observations were conducted with the VLA in its A-configuration and thus had significantly lower spatial resolutions than ours (beam sizes of 0.2--0.3\arcsec\ at 8.46\,GHz).  This implies that the jet components reached significantly greater angular separations than seen in our 2013 observations.  They also found the southern component to fade first, but the peak brightness detected in their images was from the northern component.

As in our 2013 data, several of the 2002 radio detections were made following a rebrightening during the soft state \citep[as identified by][]{2004ApJ...609..977G}, which lasted from April 6 through June 4.  The peak radio flux of 1.8\,mJy\,beam$^{-1}$ was detected on March 30, one week before the X-ray peak on April 6, and at the end of the hard X-ray state during the rise phase, as classified by \citet{2004ApJ...609..977G}.  Radio emission was not detected in the first soft-state observation on April 7, but the source reappeared as a resolved double between 17 and 26 days after the onset of the soft state.

\subsection{Plausible scenarios}

We see significant discrepancies between the observed behaviour of \xtenineteen\ and the standard expectations of intrinsically symmetric transient jets that are ejected from the core at the state transition, propagate outwards and then fade away \citep[e.g.][]{2004MNRAS.355.1105F}.  These discrepancies include the delayed appearance of the northern component, the strong evolution of the brightness and morphology of the southern component, and the measured proper motions and relative flux densities of the two components.  Possible scenarios that could explain the observations involve either absorption (either intrinsic or external to the source) that affects the observed flux density from one or both components, and the re-acceleration of particles downstream due to either internal or external shocks.  Assuming an intrinsically symmetric ejection, we now assess the relative merits of these different explanations.

\subsubsection{Internal self-absorption}

The delayed appearance of the northern component could be due to its slower expansion, for instance due to it remaining confined (e.g.\ by magnetic fields or external pressure) for longer. The faster expansion of the southern component would enable it to become optically thin earlier, allowing us to see it appear and brighten in our VLBA images several days before the northern component ceased to be affected by the relevant absorption mechanism; likely either synchrotron self-absorption or free-free absorption.  We consider this scenario to be
unlikely, as at no stage did \citet{2015MNRAS.451.3975C} observe an inverted radio spectrum, and we saw no major change in the radio spectral index on the appearance of the northern component.

\subsubsection{External obscuration}

External absorption due to dense ambient material, an equatorial wind or even an an obscuring torus could mask the synchrotron radiation until the receding knot had moved out to a point where the optical depth was sufficiently low.  While this could explain the apparent delayed appearance of the northern knot if that component was receding, this would be at odds with the higher measured proper motion of the northern component.  We would need to invoke some sort of anisotropy in the surrounding medium that could slow the approaching southern component.  Any such interaction with the surroundings would create a shock that would have been observed in the integrated radio light curves.

If the obscuration were due to free-free absorption, we can estimate the absorption coefficient
\citep{1988gera.book.....K} as
\begin{equation}
k_c \cong \frac{0.08235 n_e}{\nu^{2.1}T_{\rm e}^{1.35}},
\end{equation}
where $\nu$ is the frequency in GHz, $n_e$ is the electron density in units of cm$^{-3}$ and $k_c$ is in
pc$^{-1}$. If we assume the components appear when the optical depth ($k_c\ ds$)
reaches unity ($ds\sim 10$\,mas from the core) and $T_{\rm e}=10^4$~K, then at 8.4~GHz,
$n_e\approx2.7\times10^8$~cm$^{-3}$.  This is comparable in density to the solar corona \citep{Doschek97,Warren09}, but would need to stretch over tens of AU, and would have a total mass of $10^{-5}$--$10^{-3}M_{\odot}$ (depending on the geometry).  This seems unlikely, since the binary companion is suggested to be a main sequence star \citep{2006MNRAS.365.1387C}.  Thus, we do not favour this explanation.

\subsubsection{Internal shocks}

The components could instead be internal shocks that formed within the jets and propagated downstream \citep[e.g.][]{2004MNRAS.355.1105F,2010MNRAS.401..394J,2000A&A...356..975K}.  However, to explain the higher proper motion, lower flux density, and delayed appearance of the northern component, we would again need to invoke asymmetries in the approaching and receding jets, making this explanation similarly unsatisfactory.

\subsubsection{External shocks}

The one explanation that could naturally explain the asymmetries in the northern and southern components is that we are observing the shocks created as the jets encounter a dense region of the ISM \citep[possibly after reaching the edge of a cavity inflated by jets or accretion disc winds; see, e.g.][]{Hao09}.  Different ISM densities on either side of the source could give rise to the asymmetry, with shock-acceleration of particles at the interface then generating the observed radio emission.  

The only previous cases in which lateral expansion of jet knots have been directly observed are the BH XRB XTE J1752$-$223 and the neutron star X-ray binary Sco X-1.  In XTE J1752$-$223, \citet{Yang10} found one of the ejecta to have expanded at $0.9\pm0.1$\,mas\,d$^{-1}$, corresponding to $0.02c$ at the claimed distance of 3.5\,kpc \citep{Shaposhnikov10}.  The ejecta were also seen to be decelerating, and were inferred to be interacting with the surrounding environment on scales of a few hundred mas (thousands of AU) downstream \citep{Yang10,2011MNRAS.415..306M}.

\citet{2001ApJ...558..283F} made intensive VLBI observations of Sco X-1, resolving the system into a central core and two downstream lobes that moved outwards in opposite directions at speeds varying from $0.32$--$0.58c$.  They found that each individual knot moved ballistically for many hours, although different knots had different inferred intrinsic speeds.  During one epoch in 1998 February, they measured the expansion of the northeastern lobe, which was extended perpendicular to the jet direction and increased in size by a factor of 4.5 over 50\,min, implying an expansion velocity close to $c$.  \citet{2001ApJ...558..283F} modelled the lobes as the working surfaces where the relativistic jets of Sco X-1 impacted the ambient medium.  The electrons accelerated at the working surface then diffuse outwards along the radial direction at $0.57c$ to form an extended lobe.

A similar interaction scenario could explain much of the behaviour seen in \xtenineteen.  The late-time radio flare corresponding to the appearance and evolution of the VLBI components could be explained as the time taken for the jets launched close to the black hole to propagate outwards and impact the ISM.  The jets would have been launched at the hard-to-soft X-ray state transition (possibly giving rise to the short-lived radio flare on MJD 56601), and then decayed via adiabatic expansion to give the observed radio quenching.  When the ejected material then ran into the external medium, it would have produced a shock at the working surface, observed as the VLBI ejecta that we monitored.  The delayed appearance of the northern component would then be due to the different environmental densities on the two sides of the source.

The initial shock compression of the magnetic field at the working surface would give rise to the high degree of linear polarization seen on MJD 56605.0 (up to 13\% at 7.45\,GHz and 10\% at 5.25\,GHz).  The fractional polarization then drops to 1--2\% on MJD 56607.0 as the electrons accelerated at the shock diffuse outwards and the magnetic field becomes less ordered, before recovering to 5\% the following day as the northern component appears, giving rise to new shock compression.  The gradual rotation of the EVPA between MJD 56604 and 56607 would be due to the diffusion of the accelerated electrons away from the shock-compressed magnetic field at the working surface.  The shock acceleration at the working surface as the core of the southern component rebrightens on MJD 56606.8 could also explain the slight flattening of the radio spectrum seen between MJD 56605 and 56607, although the dominance of the halo component means that the impact of this new shock acceleration is only mild.

In this scenario, the brightness of the two components would be a reflection of both any intrinsic Doppler boosting and also the strength of the interaction with the ambient medium.  Similarly, the hotspot advance speed would be dictated by the pressures in the shocked and unshocked regions \citep{2001ApJ...558..283F}, rather than by the intrinsic speed of the jets launched close to the black hole.  Indeed, \citet{2001ApJ...558..283F} found a case where one of the hotspots clearly decelerated as it underwent a flare, due to forming a shock on interaction with the surrounding medium.  Given the potential role of the environment in giving rise to the observed differences between the northern and southern components in \xtenineteen, the constraint on $\beta\cos\theta$ derived from the proper motions of the northern and southern components {\it may not be representative of the underlying, intrinsic jet parameters}.  Similarly, the derived core location may not be correct.  Nonetheless, if we assume that the original ejection event was on or before the detection of the short-lived radio flare detected by AMI-LA, we still infer a low projected jet speed.  The first detections of the northern and southern components are separated by 25\,mas and four days, implying a maximum projected jet speed of $0.29c (d/8{\rm\,kpc})$. 

Under this scenario, it is interesting to consider the 40-ks NuSTAR flare reported by \citet{Zhang15} and \citet{Tao15} (see Section~\ref{sec:xray}).  If this was indeed a flare from the corona, there is no corresponding signature in the VLBI images until either the rebrightening of the core of the southern component on MJD 56606.8, or the development of the northern component on MJD 56607.8.  While these associations with downstream events are tentative at best, we can determine the implications of such a scenario, which would then give a time difference of 1.5--2.5 days between the X-ray flare and any radio signature.  The southern core on MJD 56606.8 is separated from the inferred binary location by $12.5\pm3.3$\,mas, and the northern component on MJD 56607.8 by $18.9\pm3.6$\,mas.  This would therefore imply proper motions of $\sim8\pm2$\,mas\,d if the energy were to propagate directly downstream from the core to the working surfaces, for a projected speed (in the plane of the sky) of $0.3$--$0.4c$.  This would be similar to the scenario observed in Sco X-1, where the energy was transferred from the core to the lobes at $>0.95c$, much faster than the advance speed of the hotspot \citep{Fomalont01}.  

\subsection{Energetics}

The southern component reached its peak brightness on MJD 56606.8, when it had a total integrated flux of $15.6\pm0.8$\,mJy, and a Gaussian full-width at half maximum (FWHM$=2\sqrt{2\ln 2}\sigma$) of 14\,mas. Assuming a representative distance of 8\,kpc and a spherical geometry, this would imply an emitting volume of $2\times10^{45}$\,cm$^{-3}$. We performed a standard minimum energy calculation \citep[e.g.][]{1970ranp.book.....P,1994hea..book.....L,1996tra..book.....R} by assuming an underlying electron energy distribution index $p=2.2$ (corresponding to a spectral index of $-0.6$ when fully optically thin), a spectrum extending between at least $10^7$ and $10^{11}$\,Hz, and as much energy in protons as there is in relativistic electrons. The minimum energy and corresponding minimum magnetic field strength were determined via

\begin{equation}
E_{\rm{min}}=\frac{7}{6\mu_0}V^{3/7}\left[\frac{3\mu_0}{2}G(\alpha)\eta L_{\nu}\right]^{4/7},
\end{equation}
\begin{equation}
B_{\rm{min}}=\left[\frac{3\mu_0}{2}\frac{G(\alpha)\eta L_{\nu}}{V}\right]^{2/7},
\end{equation}
where $G(\alpha)$ is a constant that loosely depends on
$\alpha,\nu_{\rm{min}}$ and $\nu_{\rm{max}}$. We find a minimum energy of $E_{\rm min}=5\times10^{41}$\,erg, and a minimum energy field of $B_{\rm min} = 50$\,mG.

In comparison, both \grs\ and XTE J1748$-$288 were found to have had ejection events with minimum energies of a few times $10^{42}$\,erg \citep{1999MNRAS.304..865F,Brocksopp07}.  In contrast, a detailed study of a sequence of ejection events in XTE J1752$-$223 showed minimum energies for the different events ranging from $10^{41}$--$7\times10^{42}$\,erg, showing that different ejection events in an individual source can vary greatly in energy \citep{Brocksopp13}. Thus, the minimum energy derived for the main radio flare observed in \xtenineteen\ was on the low end of what has been seen from other BH XRB jets, but by no means an outlier relative to the overall distribution.

Knowing the distance downstream at which the jets slow down and form hotspots, we can use the formalism of \citet{2002A&A...388L..40H} to constrain the environmental density, which is given by
\begin{equation}
n \sim \frac{E_{44}}{\Gamma_5^2 \theta_5^2 d_{16}^3}{\rm\,cm}^{-3},
\label{eq:energy}
\end{equation}
where $E_{44}$ is the energy in the jets in units of $10^{44}$\,erg, the jet Lorentz factor is $\Gamma=5\Gamma_5$, the knot opening angle is $\theta=5^{\circ}\theta_5$, and the jet decelerates due to its interaction with the surroundings at a distance $10^{16}d_{16}$\,cm.  We assume a canonical Lorentz factor of $\Gamma=2$, an opening angle of $\theta=58^{\circ}$ (as derived from the size and downstream distance of the southern lobe on MJD 56606.8) and a distance downstream from the binary system of $d=0.15\times10^{16}$\,cm. This implies a density $n=13 E_{44}$\,cm$^{-3}$.  Given the minimum energy determined above, this would place a lower limit on the density of $n\geq0.07$\,cm$^{-3}$, such that the jets would not need to be far out of equipartition in order for the local ISM density to be comparable to the canonical value of 1\,cm$^{-3}$.  If (as seems likely) the actual jet opening angle is smaller (i.e.\ the measured size instead reflects the lateral expansion of the lobe, and not of the jet itself), the inferred density would be higher still.  This is in stark contrast to \grs\ and \gro, which \citet{2002A&A...388L..40H} found to be located in significantly more underdense environments, with $n<10^{-3}$\,cm$^{-3}$.  

\citet{Hao09} modelled the jet-ISM interactions in XTE J1550$-$564 and H1743$-$322 by assuming that the observed emission was produced at the reverse shock generated when the jets, having propagated outwards through a low-density cavity, encountered the denser, surrounding ISM.  They inferred similarly low ISM densities of $\sim3\times10^{-3}$\,cm$^{-3}$ around H1743$-$322 and the eastern jet of XTE J1550$-$564. However, the latter source was found to lie in a very inhomogeneous environment, with the western jet running into a higher-density medium with $n=0.12$\,cm$^{-3}$.  Without more detailed knowledge of the core location or the ejection time (which we inferred in Section~\ref{sec:propermotions} by assuming ballistic motions for our components and extrapolating back to zero separation), we do not have sufficient constraints to apply such a model to \xtenineteen. 
However, should our target be located in a low-density cavity \citep[as hypothesised to exist around all microquasar systems by][]{Hao09}, the extent of the cavity must be significantly lower than in XTE J1550$-$564 or H1743$-$322, given that the jet-ISM interaction was seen to occur on scales of a few tens of milliarcseconds rather than several arcseconds.  This would imply that the jets or accretion disc winds responsible for inflating the cavity were significantly weaker in \xtenineteen\ (and also in XTE J1752$-$223, where jet-ISM interactions were seen on scales of a few hundred milliarcseconds), or that the outflows responsible for creating the cavity had been active for less time.

The different distances to which the jets propagated during the 2002 and 2013 outbursts would seem to argue against a single, unchanging, large-scale cavity around \xtenineteen, as the jet-ISM interactions occurred much closer to the central binary system in 2013 than in 2002.  Nonetheless, the 2002 outburst also seems consistent with an external shock scenario, with the system going through an initial flare, followed by a period of non-detections, and finally a rebrightening during which time the resolved ejecta were detected (see Section~\ref{sec:2002}).  As in 2013, the downstream rebrightening of these resolved ejecta would seem to argue for interactions between the jet and the ISM.  Given the relatively sparse time sampling, we cannot determine whether the southern interaction region ever got as bright as its northern counterpart prior to April 23 (the date of the image in Fig.~\ref{fig:2002}).  Regardless, the appearance of the knots further downstream in 2002 suggests a difference in the properties of either the jets or the environment.

In the absence of a large-scale cavity, we can instead explore the uniform density scenario of \citet{2002A&A...388L..40H}.  Had there been no significant change in the external density between 2002 and 2013, Equation~(\ref{eq:energy}) would suggest that despite comparable peak X-ray fluxes \citep{2015MNRAS.451.3975C}, the jets of \xtenineteen\ were more energetic in 2002 than in 2013, such that they could propagate out to a larger distance. With the two components in Fig.~\ref{fig:2002} being separated by $326\pm35$\,mas, the most distant component must be at least 163\,mas from the core, implying a downstream distance $d_{16}=2.0$.  Therefore, if the environmental density, jet Lorentz factor and opening angle were unchanged, $E_{44}$ would need to have been 2000 times as great in 2002 as in 2013.  Given the comparable peak X-ray fluxes, this seems implausible.  Either the properties of the environment were different (e.g.\ a decrease in the density on moving further away from the binary system, since Equation~(\ref{eq:energy}) assumes a uniform density), or the jet Lorentz factor and/or opening angle were smaller in 2002.

\subsection{Comparison to other systems}

Laterally expanding jets are a rare phenomenon in low-mass BH XRBs \citep{2006MNRAS.367.1432M}, with the only example to date having been seen in XTE J1752$-$223 \citep{Yang10}, whose jet was also thought to have been interacting with the ISM.  Similar expanding working surfaces were also seen in the neutron star XRB Sco X-1 \citep{2001ApJ...558..283F}.  These were regenerated during multiple individual ejection events, always moving ballistically, and appearing at the same position angle with respect to the central source, in the same way that \xtenineteen\ showed jets along the same position angle in both 2002 and 2013.  However, the working surfaces moved outwards faster in both XTE J1752$-$223 and Sco X-1 than in \xtenineteen\ (despite the lower calculated minimum energy of $6\times10^{39}$\,erg in Sco X-1).

A third system showing laterally-expanding outflows was the accreting source CI Cam during its 1998 outburst.  VLBI observations by \citet{2004ApJ...615..432M} suggested that the compact object ejected strong jets that were smothered by a dense circumstellar medium, leading to a much less collimated outflow than typically observed from X-ray binary jets.  However, the recent Gaia distance measurement to this system has suggested that it is an accreting white dwarf rather than a neutron star XRB \citep{Wijngaarden16, Barsukova06}.

While lateral expansion appears to be rare, the deceleration of jets as they interact with the ISM has been seen in a number of sources in recent years, over a wide range of scales.  The jets in XTE J1752$-$223 decelerated on a scale of a few hundred mas \citep{Yang10,2011MNRAS.418L..25Y,2011MNRAS.415..306M}, on a timescale of a few tens of days following the flare.  On the other hand, the jets of XTE J1550$-$564 and H1743$-$322 were both seen to decelerate significantly further downstream, at distances of several arcseconds, and hundreds of days following the original outburst \citep{2002Sci...298..196C,Kaaret03,2005ApJ...632..504C}.  According to the formalism of \citet{2002A&A...388L..40H}, this would imply either smaller opening angles, lower environmental densities, or more energetic jets in the latter two cases.  Alternatively, should these systems in fact be surrounded by low-density cavities inflated by accretion disc winds or jets \citep[as suggested by][]{Hao09}, the cavities must be significantly smaller, and hence the strength of the outflows correspondingly weaker in XTE J1752$-$223 and \xtenineteen.

\section{Conclusions}
\label{section:conclusions}

We have presented a rare example of lateral expansion of the jets from a BH XRB. Our high-resolution VLBI monitoring of the 2013 outburst of \xtenineteen\ has shown the development of asymmetric, resolved, expanding ejecta following a hard to soft state transition. These ejecta move in opposite directions with proper motions of
$2-3$~mas~day$^{-1}$, implying relatively low jet speeds of $<0.3c(d/8{\rm\,kpc}$). The ejecta appear to expand isotropically at a rate of $\sim3$~mas~day$^{-1}$. We interpret these jet components as the working surfaces where the jets ejected at the hard-to-soft state transition impacted the ISM, creating a moving shock front that accelerated particles, which subsequently diffused outwards over time.  Minimum energy calculations showed that the jet energy was at the low end of the observed distribution of BH XRB jets, and suggested that the surrounding ISM was denser than seen around other systems such as \grs\ and \gro.

Comparison with the photometric and polarimetric data from integrated radio light curves shows that the jet-ISM interaction created a long-lived radio flare, whose evolution was linked to the appearance and evolution of the two resolved VLBI components.  This long-lived flare reached flux densities comparable to the rapid initial flare detected by AMI-LA at the state transition, and suggests that care should be taken in conducting jet-disc coupling studies without accompanying high angular resolution images.  Integrated light curves alone are unable to distinguish between internal and external shock scenarios, such that VLBI monitoring is required to accurately interpret contemporaneous radio and X-ray phenomenology.

\section*{Acknowledgments}

We are deeply indebted to Peter Curran, who drove much of this work and was responsible for a significant amount of the data collation and interpretation before his death.  His contributions to the field will be greatly missed by his many colleagues.  The National Radio Astronomy Observatory is a facility of the National Science Foundation operated under cooperative agreement by Associated Universities, Inc. The European VLBI Network is a joint facility of independent European, African, Asian, and North American radio astronomy institutes. Scientific results from data presented in this publication are derived from the following EVN project codes: RR007 and RR009. This work made use of the Swinburne University of Technology software correlator, developed as part of the Australian Major National Research Facilities Programme and operated under licence. APR acknowledges funding via an EU Marie Curie Intra-European Fellowship under contract no. 2012-331977. The work was also supported by ERC grant 267697 ``4~PI~SKY: Extreme Astrophysics with Revolutionary Radio Telescopes''. JCAMJ is supported by an Australian Research Council Future Fellowship (FT140101082). This work was supported by Australian Research Council grant DP120102393. DA acknowledges support from the Royal Society.

\label{lastpage}


\begin{thebibliography}{9}

\bibitem[\protect\citeauthoryear{Brocksopp et al.}{2007}]{Brocksopp07}
Brocksopp C., Miller-Jones J.~C.~A., Fender R.~P., Stappers B.~W., 2007, MNRAS, 378, 1111

\bibitem[\protect\citeauthoryear{Brocksopp et al.}{2013}]{Brocksopp13}
Brocksopp C., Corbel S., Tzioumis A., Broderick J.~W., Rodriguez J., Yang J., Fender R.~P., Paragi Z., 2013, MNRAS, 432, 931

\bibitem[\protect\citeauthoryear{Barsukova et al.}{2006}]{Barsukova06}
Barsukova E.~A., Borisov N.~V., Burenkov A.~N., Goranskii V.~P., Klochkova V.~G., Metlova N.~V., 2006, Astronomy Reports, 50, 664

\bibitem[Belloni(2010)]{2010LNP...794...53B} Belloni, T.~M.\ 2010, Lecture Notes in Physics, Berlin Springer Verlag, 794, 53 

\bibitem[Chaty et al.(2002)]{2002MNRAS.337L..23C} Chaty, S., Mignani, 
R.~P., \& Israel, G.~L.\ 2002, \mnras, 337, L23 

\bibitem[Chaty et al.(2006)]{2006MNRAS.365.1387C} Chaty, S., Mignani, 
R.~P., \& Israel, G.~L.\ 2006, \mnras, 365, 1387 

\bibitem[Corbel et al.(2002)]{2002Sci...298..196C} Corbel S., Fender R.~P., Tzioumis A.~K., Tomsick J.~A., Orosz J.~A., Miller J.~M., Wijnands R., Kaaret P.\ 2002, Science, 298, 196

\bibitem[Corbel et al.(2003)]{2003NewAR..47..477C} Corbel S., Fender R.~P., Tzioumis A.~K., Tomsick J.~A., Orosz J.~A., Miller J.~M., Wijnands R., Kaaret P.\ 2003, \nar, 47, 477 

\bibitem[Corbel et al.(2005)]{2005ApJ...632..504C} Corbel S., Kaaret P., Fender R.~P., Tzioumis A.~K., Tomsick J.~A., Orosz J.~A.\ 2005, \apj, 632, 504 

\bibitem[Corbel et al.(2013)]{2013MNRAS.431L.107C} Corbel, S.\ et al.\ 2013, \mnras, 431, L107 

\bibitem[\protect\citeauthoryear{Coriat et al.}{2013}]{Coriat13}
Coriat M., Tzioumis T., Corbel S., Fender R., 2013, The Astronomer's Telegram, 5575, 1

\bibitem[\protect\citeauthoryear{Curran et al.}{2014}]{2014MNRAS.437.3265C} Curran P.~A., et al., 2014, MNRAS, 437, 3265 

\bibitem[Curran et al.(2015)]{2015MNRAS.451.3975C} Curran, P.~A.\ et al.\ 2015, \mnras, 451, 3975 

\bibitem[Deller et al.(2007)]{2007PASP..119..318D}
Deller A.~T., Tingay S.~J., Bailes M., West C.\ 2007, \pasp, 119, 318

\bibitem[Dhawan et al.(2000)]{2000ApJ...543..373D} Dhawan, V., Mirabel, 
I.~F., \& Rodr{\'{\i}}guez, L.~F.\ 2000, \apj, 543, 373 

\bibitem[\protect\citeauthoryear{Doschek et al.}{1997}]{Doschek97}
Doschek G.~A., Warren H.~P., Laming J.~M., Mariska J.~T., Wilhelm K., Lemaire P., Sch{\"u e}, U., Moran T.~G., 1997, ApJ, 482, L109

\bibitem[Fender et al.(1999)]{1999MNRAS.304..865F} Fender R.~P., Garrington S.~T., McKay D.~J., Muxlow T.~W.~B., Pooley G.~G., Spencer R.~E., Stirling A.~M., Waltman E.~B.\ 1999, \mnras, 304, 865 

\bibitem[Fender(2006)]{2006csxs.book..381F} Fender, R.\ 2006, in Lewin W. H. G., van der Klis M., eds, Cambridge Astrophysics Series, No. 39, Compact Stellar X-ray Sources. Cambridge Univ. Press, Cambridge, p. 381

\bibitem[Fender et al.(2004)]{2004MNRAS.355.1105F} Fender, R., Belloni, T., \& Gallo, E.\ 2004, \mnras, 355, 1105

\bibitem[\protect\citeauthoryear{Fender, Homan \& Belloni}{Fender et al.}{2009}]{Fender09}
Fender R.~P., Homan J., Belloni T.~M., 2009, MNRAS, 396, 1370

\bibitem[Feroci et al.(2002)]{2002IAUC.7861....2F} Feroci, M., Reboa, L., 
\& BEPPOSAX Team 2002, \iaucirc, 7861, 2 

\bibitem[Fomalont et al.(2001)]{2001ApJ...558..283F} Fomalont, E.~B., 
Geldzahler, B.~J., \& Bradshaw, C.~F.\ 2001, \apj, 558, 283 

\bibitem[\protect\citeauthoryear{Fomalont, Geldzahler \& Bradshaw}{Fomalont et al.}{2001}]{Fomalont01}
Fomalont E.~B., Geldzahler B.~J., Bradshaw C.~F., 2001, ApJ, 553, L27

\bibitem[Gallo et al.(2005)]{2005Natur.436..819G} Gallo E., Fender R., Kaiser C., Russell D., Morganti R., Oosterloo T., Heinz S.\ 2005, \nat, 436, 819 

\bibitem[Garnavich et al.(2002)]{2002IAUC.7877....4G} Garnavich, P., Quinn, 
J., \& Callanan, P.\ 2002, \iaucirc, 7877, 4 

\bibitem[G{\"o}{\v g}{\"u}{\c s} et al.(2004)]{2004ApJ...609..977G}
G{\"o}{\v g}{\"u}{\c s}, E. et al., 2004, ApJ, 609, 977

\bibitem[Greisen(2003)]{2003ASSL..285..109G} Greisen E. W., 2003, in Heck A., ed., Information Handling in Astronomy: Historical Vistas. Kluwer, Dordrecht, p. 109

\bibitem[\protect\citeauthoryear{Hao \& Zhang}{2009}]{Hao09}
Hao J.~F., Zhang S.~N., 2009, ApJ, 702, 1648

\bibitem[Heinz(2002)]{2002A&A...388L..40H} Heinz, S.\ 2002, \aap, 388, L40 

\bibitem[Heinz et al.(2008)]{2008ApJ...686.1145H} Heinz, S., Grimm, H.~J., Sunyaev, R.~A., \& Fender, R.~P.\ 2008, \apj, 686, 1145-1154 

\bibitem[Hjellming 
\& Rupen(1995)]{1995Natur.375..464H} Hjellming, R.~M., \& Rupen, M.~P.\ 1995, \nat, 375, 464 

\bibitem[in 't Zand et al.(2002a)]{2002IAUC.7873....1I} in 't Zand, 
J.~J.~M., Capalbi, M., \& Perri, M.\ 2002a, \iaucirc, 7873, 1 

\bibitem[in't Zand et 
al.(2002b)]{2002A&A...394..553I} in't Zand, J.~J.~M., Miller, J.~M., Oosterbroek, T., \& Parmar, A.~N.\ 2002b, \aap, 394, 553 

\bibitem[Jamil et al.(2010)]{2010MNRAS.401..394J} Jamil, O., Fender, R.~P., 
\& Kaiser, C.~R.\ 2010, \mnras, 401, 394 

\bibitem[Jonker et al.(2004)]{2004MNRAS.351.1359J} Jonker P.~G., Gallo E., Dhawan V., Rupen M., Fender R.~P., Dubus G.\ 2004, \mnras, 351, 1359 

\bibitem[\protect\citeauthoryear{Kaaret et al.}{2003}]{Kaaret03}
Kaaret P., Corbel S., Tomsick J.~A., Fender R., Miller J.~M., Orosz J.~A., Tzioumis A.~K., Wijnands R., 2003, ApJ, 582, 945

\bibitem[Kaiser et al.(2000)]{2000A&A...356..975K} Kaiser, C.~R., Sunyaev, R., \& Spruit, H.~C.\ 2000, \aap, 356, 975 

\bibitem[\protect\citeauthoryear{Kalemci et al.}{2013}]{Kalemci13}
Kalemci E., Din{\c c }, T., Tomsick J.~A., Buxton M.~M., Bailyn C.~D., Chun Y.~Y., 2013, ApJ, 779, 95

\bibitem[Kaiser(2006)]{2006MNRAS.367.1083K} Kaiser, C.~R.\ 2006, \mnras, 
367, 1083 

\bibitem[Keimpema et al.(2015)]{2015ExA....39..259K} Keimpema, A., Kettenis, M.~M., Pogrebenko, S.~V., et al.\ 2015, Experimental Astronomy, 39, 259 

\bibitem[Kettenis et al.(2006)]{2006ASPC..351..497K} Kettenis, M., van Langevelde, H.~J., Reynolds, C., \& Cotton, B.\ 2006, in Gabriel C., Arviset C., Ponz D., Solano E., eds, ASP Conf. Ser. Vol. 351, Astronomical Data Analysis Software and Systems XV. Astron. Soc. Pac., San Francisco, p. 497

\bibitem[Krimm et al.(2013a)]{2013ATel.5523....1K} Krimm, H.~A.,\ et al.\ 2013, The Astronomer's Telegram, 5523, 1 

\bibitem[Krimm et al.(2013b)]{2013ATel.5529....1K} Krimm, H.~A., Kennea, 
J.~A., \& Holland, S.~T.\ 2013, The Astronomer's Telegram, 5529, 1 

\bibitem[\protect\citeauthoryear{Kubota, Makishima \& Ebisawa}{Kubota et al.}{2001}]{Kubota01}
Kubota A., Makishima K., Ebisawa K., 2001, ApJ, 560, L147

\bibitem[Ling et al.(2009)]{2009ApJ...695.1111L} Ling Z., Zhang S.~N., Tang S., 2009, ApJ, 695, 1111

\bibitem[Longair(1994)]{1994hea..book.....L} Longair, M.~S.\ 1994, 
High energy astrophysics. Vol.2: Stars, the galaxy and the interstellar medium, 2nd Ed. Cambridge: Cambridge University Press  

\bibitem[\protect\citeauthoryear{Maccarone}{2003}]{Maccarone03}
Maccarone T.~J., 2003, A\&A, 409, 697

\bibitem[McHardy et al.(2006)]{2006Natur.444..730M} McHardy, I.~M., 
Koerding, E., Knigge, C., Uttley, P., 
\& Fender, R.~P.\ 2006, \nat, 444, 730 

\bibitem[Miller et al.(2009)]{Miller09}
Miller J.~M., Reynolds C.~S., Fabian A.~C., Miniutti G., Gallo L.~C., 2009, ApJ, 697, 900

\bibitem[Miller-Jones et al.(2004)]{2004ApJ...600..368M}
Miller-Jones J.~C.~A., Blundell K.~M., Rupen M.~P., Mioduszewski A.~J., Duffy P., Beasley A.~J.\ 2004, \apj, 600, 368

\bibitem[\protect\citeauthoryear{Miller-Jones, Blundell, \& Duffy}{2004b}]{2004ApJ...603L..21M} Miller-Jones J.~C.~A., Blundell K.~M., Duffy P., 2004b, ApJ, 603, L21 

\bibitem[Miller-Jones et al.(2006)]{2006MNRAS.367.1432M} Miller-Jones, 
J.~C.~A., Fender, R.~P., \& Nakar, E.\ 2006, \mnras, 367, 1432 

\bibitem[Miller-Jones et al.(2007)]{2007MNRAS.375.1087M} Miller-Jones J.~C.~A., Rupen M.~P., Fender R.~P., Rushton A., Pooley G.~G., Spencer R.~E.\ 2007, \mnras, 375, 1087 

\bibitem[Miller-Jones et al.(2011)]{2011MNRAS.415..306M} Miller-Jones J.~C.~A., Jonker P.~G., Ratti E.~M., Torres M.~A.~P., Brocksopp C., Yang J., Morrell N.~I.\ 2011, \mnras, 415, 306

\bibitem[Miller-Jones et al.(2012)]{2012MNRAS.421..468M} Miller-Jones J.~C.~A. et al., 2012, \mnras, 421, 468

\bibitem[\protect\citeauthoryear{Miller-Jones, Sivakoff \& Krimm}{Miller-Jones et al.}{2013}]{Miller-Jones13}
Miller-Jones J.~C.~A., Sivakoff G.~R., Krimm H.~A., 2013, The Astronomer's Telegram, 5530, 1

\bibitem[Mioduszewski et al.(2001)]{2001ApJ...553..766M}
Mioduszewski A.~J., Rupen M.~P., Hjellming R.~M., Pooley G.~G., Waltman E.~B.\ 2001, \apj, 553, 766

\bibitem[Mioduszewski \& Rupen(2004)]{2004ApJ...615..432M} Mioduszewski, A.~J., \& Rupen, M.~P.\ 2004, \apj, 615, 432 

\bibitem[Mirabel et al.(1992)]{1992Natur.358..215M} Mirabel, I.~F., Rodriguez, L.~F., Cordier, B., Paul, J., \& Lebrun, F.\ 1992, \nat, 358, 215 

\bibitem[Mirabel 
\& Rodr{\'{\i}}guez(1994)]{1994Natur.371...46M} Mirabel, I.~F., \& Rodr{\'{\i}}guez, L.~F.\ 1994, \nat, 371, 46 

\bibitem[Mirabel 
\& Rodriguez(2002)]{2002S&T...103e..32M} Mirabel, I.~F., \& Rodriguez, L.~F.\ 2002, \skytel, 103, 32 

\bibitem[Mirabel et al.(2011)]{2011A&A...528A.149M} Mirabel, I.~F., Dijkstra, M., Laurent, P., Loeb, A., \& Pritchard, J.~R.\ 2011, \aap, 528, AA149 

\bibitem[Pacholczyk(1970)]{1970ranp.book.....P} Pacholczyk A. G., 1970, Radio Astrophysics. Freeman \& Co., San Francisco

\bibitem[Pakull et al.(2010)]{2010Natur.466..209P} Pakull, M.~W., Soria, R., \& Motch, C.\ 2010, \nat, 466, 209 

\bibitem[Reid et al.(2014)]{2014ApJ...796....2R} Reid M.~J., McClintock J.~E., Steiner J.~F., Steeghs D., Remillard R.~A., Dhawan V., Narayan R.\ 2014, \apj, 796, 2

\bibitem[Reynolds et al.(2002)]{2002astro.ph..5118R} Reynolds, C., Paragi, Z., \& Garrett, M.\ 2002, arXiv:astro-ph/0205118 

\bibitem[Rodriguez et al.(1992)]{1992ApJ...401L..15R} Rodriguez, L.~F., Mirabel, I.~F., \& Marti, J.\ 1992, \apjl, 401, L15 

\bibitem[Rohlfs \& Wilson(1996)]{1996tra..book.....R} Rohlfs K., Wilson T. L., 1996, Tools of Radio Astronomy (Astronomy and Astrophysics Library). Springer-Verlag, Berlin, p.190 

\bibitem[Rupen et al.(2002a)]{2002IAUC.7874....1R} Rupen, M.~P., Dhawan, V., 
\& Mioduszewski, A.~J.\ 2002a, \iaucirc, 7874, 1 

\bibitem[Rupen et al.(2002b)]{2002IAUC.8029....2R} Rupen, M.~P., 
Mioduszewski, A.~J., \& Dhawan, V.\ 2002b, \iaucirc, 8029, 2 

\bibitem[Rushton et al.(2010)]{2010MNRAS.401.2611R} Rushton, A., Spencer, 
R.~E., Pooley, G., \& Trushkin, S.\ 2010, \mnras, 401, 2611 

\bibitem[Rushton et 
al.(2010)]{2010A&A...524A..29R} Rushton, A., Spencer, R., Fender, R., \& Pooley, G.\ 2010, \aap, 524, A29 

\bibitem[\protect\citeauthoryear{Rushton et al.}{2013a}]{Rushton13a}
Rushton A.~P., Fender R., Anderson G., Staley T., Rumsey C., Titterington D., 2013a, The Astronomer's Telegram, 5532, 1

\bibitem[\protect\citeauthoryear{Rushton et al.}{2013b}]{Rushton13b}
Rushton A.~P., Fender R., Anderson G., Staley T., Rumsey C., Titterington D., 2013b, The Astronomer's Telegram, 5551, 1

\bibitem[Russell et al.(2007)]{2007MNRAS.376.1341R} Russell, D.~M., Fender, R.~P., Gallo, E., \& Kaiser, C.~R.\ 2007, \mnras, 376, 1341 

\bibitem[Sell et al.(2015)]{2015MNRAS.446.3579S} Sell, P.~H., Heinz, S., Richards, E., et al.\ 2015, \mnras, 446, 3579 

\bibitem[\protect\citeauthoryear{Shaposhnikov et al.}{2010}]{Shaposhnikov10}
Shaposhnikov N., Markwardt C., Swank J., Krimm H., 2010, ApJ, 723, 1817

\bibitem[\protect\citeauthoryear{Shepherd}{1997}]{Shepherd97}
Shepherd M.~C., in Hunt G., Payne H. E., eds, ASP Conf. Ser.
Vol. 125, Astronomical Data Analysis Software and Systems VI. Astron. Soc. Pac., San Francisco, p. 77

\bibitem[Stirling et al.(2001)]{2001MNRAS.327.1273S} Stirling A.~M., Spencer R.~E., de la Force C.~J., Garrett M.~A., Fender R.~P., Ogley R.~N.\ 2001, \mnras, 327, 1273 

\bibitem[Szomoru(2008)]{2008evn..confE..40S} Szomoru, A.\ 2008, The role of VLBI in the Golden Age for Radio Astronomy, 40 

\bibitem[Tao et al.(2015)]{Tao15}
Tao L. et al., 2015, ApJ, 811, 51

\bibitem[Tingay et al.(1995)]{1995Natur.374..141T}
Tingay S.~J. et al., 1995, \nat, 374, 141

\bibitem[Verschuur \& Kellermann(1988)]{1988gera.book.....K} Verschuur G. L., Kellermann K. I., eds, 1988, Galactic and Extragalactic Radio Astronomy, 2nd edn. Springer-Verlag, New York 

\bibitem[Wagner \& Starrfield(2002)]{2002ATel...86....1W} Wagner, R.~M., \& Starrfield, S.\ 2002, The Astronomer's Telegram, 86, 1 

\bibitem[\protect\citeauthoryear{Warren \& Brooks}{2009}]{Warren09}
Warren H.~P., Brooks D.~H., 2009, ApJ, 700, 762

\bibitem[Wiersema et al.(2009)]{2009MNRAS.397L...6W} Wiersema, K., Russell, D.~M., Degenaar, N., et al.\ 2009, \mnras, 397, L6 

\bibitem[\protect\citeauthoryear{Wijngaarden et al.}{2016}]{Wijngaarden16}
Wijngaarden M.~J.~P., Gourdji K., Oostrum L.~C., Henrichs H.~F., 2016, The Astronomer's Telegram, 9634, 1

\bibitem[Woods et al.(2002)]{2002IAUC.7856....1W} Woods P.~M., Kouveliotou C., Finger M.~H., Gogus E., Swank J., Markwardt C., Strohmayer T.\ 2002, \iaucirc, 7856, 1 

\bibitem[\protect\citeauthoryear{Yang et al.}{2010}]{Yang10}
Yang J., Brocksopp C., Corbel S., Paragi Z., Tzioumis T., Fender R.~P., 2010, MNRAS, 409, L64

\bibitem[Yang et al.(2011)]{2011MNRAS.418L..25Y} Yang J., Paragi Z., Corbel S., Gurvits L.~I., Campbell R.~M., Brocksopp C.\ 2011, \mnras, 418, L25 

\bibitem[Yoon et al.(2011)]{2011ApJ...742...25Y} Yoon, D., Morsony, B., Heinz, S., et al.\ 2011, \apj, 742, 25 

\bibitem[Zhang et al.(2015)]{Zhang15}
Zhang L., Chen L., Qu J.-l., Bu Q.-c., Zhang W., 2015, ApJ, 813, 90

\bibitem[Zwart et al.(2008)]{2008MNRAS.391.1545Z} Zwart, J.~T.~L.\ et al.\ 2008, \mnras, 391, 1545 

\end{thebibliography}
\end{document}